\documentclass[a4paper, 10pt, onecolumn, accepted=2025-11-27]{quantumarticle}
\pdfoutput=1

\usepackage{amscd, amsfonts, amsmath, amssymb, amsthm}
\usepackage{bbold}
\usepackage{braket}
\usepackage{cite}
\usepackage{color}
\usepackage{comment}
\usepackage{dsfont}
\usepackage{epsfig}
\usepackage{graphics}
\usepackage{graphicx}
\DeclareGraphicsExtensions{bmp, eps}
\usepackage{hyperref} 
\usepackage{longtable}
\usepackage{makeidx}
\usepackage{multicol}
\usepackage[normalem]{ulem}
\usepackage[T1]{fontenc}

\hypersetup{
     colorlinks = true,
     citecolor  = blue,
     linkcolor  = blue,
     urlcolor   = blue 
}

\def\be{\begin{equation}}
\def\ee{\end{equation}}
\def\bea{\begin{eqnarray}}
\def\eea{\end{eqnarray}}

\newcommand{\sect}[1]{\setcounter{equation}{0}\section{#1}}

\renewcommand{\theequation}{\arabic{section}.\arabic{equation}}
\newcommand{\appA}[1]{\setcounter{equation}{0}\section*{Appendix A}}

\def\Gal{\overline{{\mathcal{G}}}}

\title{Quantum Galilei group as quantum reference frame transformations}

\author{Angel Ballesteros}
\affiliation{Departamento de F\'isica,  Universidad de Burgos,  09001 Burgos, Spain.}
\email{angelb@ubu.es}

\author{Diego Fernandez-Silvestre}
\affiliation{Departamento de Matem\'aticas y Computaci\'on,  Universidad de Burgos,  09001 Burgos, Spain.}
\email{dfsilvestre@ubu.es}

\author{Flaminia Giacomini}
\affiliation{Institute for Theoretical Physics, ETH Z{\"u}rich, 8093 Z{\"u}rich, Switzerland.}
\affiliation{Dipartimento di Fisica,
Universit{\`a} di Roma Tor Vergata,
Via della Ricerca Scientifica 1, 00133, Roma, Italy.}
\email{flaminia.giacomini@uniroma2.it}

\author{Giulia Gubitosi}
\affiliation{Dipartimento di Fisica Ettore Pancini, Universit\`a di Napoli Federico II, and INFN, Sezione di Napoli, Complesso Univ. Monte S. Angelo, I-80126 Napoli, Italy.}
\email{giulia.gubitosi@unina.it}

\begin{document}

\maketitle

\begin{abstract}
	 
Quantum groups have been widely explored as a tool to encode possible nontrivial generalisations of reference frame transformations, relevant in quantum gravity. In quantum information, it was found that the  reference frames can be associated to  quantum particles, leading to quantum reference frames transformations. The connection between these two frameworks is still unexplored, but if clarified it will lead to a more profound understanding of symmetries in quantum mechanics and quantum gravity. 

Here, we establish a correspondence between quantum reference frame transformations and transformations generated by a quantum deformation of the Galilei group with commutative time, taken at first order in the quantum deformation parameter. This is found once the quantum group noncommutative transformation parameters are represented on the phase space of a quantum particle, and upon setting the quantum deformation parameter to be proportional to the inverse of the mass of the particle serving as the quantum reference frame. These results allow us to show that quantum reference frame transformations are physically relevant when the state of the quantum reference frame is in a quantum superposition of semiclassical states. We conjecture that the all-order quantum Galilei group describes quantum reference frame transformations between more general quantum states of the quantum reference frame.

\end{abstract} 



\tableofcontents


\sect{Introduction}

In quantum mechanics, the notion of reference frame is abstracted from an idealised physical system. In this picture, reference frames are sharply defined and have no dynamics. Such an idealised notion of reference frames has been challenged for different reasons. For instance, as soon as we resort to physical considerations, we are forced to abandon idealised reference frames, because physically meaningful quantities are operationally defined by measuring them with a physical system, and taking as reference a physical object. In quantum gravity, various independent arguments point towards the necessity of replacing the classical notion of spacetime with a more general structure at the Planck scale. In such a quantum spacetime, the idealised notion of reference frames is insufficient.

Both quantum information and quantum gravity communities have, since 1967~\cite{DeWitt1967, Aharonov1967a}, pointed towards the necessity of considering ``quantum reference frames'' (QRFs), namely reference frames associated to dynamical material systems obeying the laws of quantum theory. Several approaches to QRFs have been developed in parallel in quantum gravity~\cite{DeWitt1967, Rovelli:1990pi, Kuchar:1990vy, Brown:1994py, Brown:1995fj, Dittrich:2004cb, Tambornino:2011vg} and in quantum information~\cite{Aharonov1967a, Aharonov1967b, Aharonov1984, Poulin:2006ryq, Bartlett2007, Spekkens2008, Angelo2011, Palmer:2013zza, Katz2015, Smith2016, Busch2016, Loveridge:2016tnh, Loveridge:2017pcv}, addressing different aspects of the consequences of abandoning idealised reference frames. Recently, there has been renewed interest in the topic, and a new wave of results~\cite{Giacomini:2017zju, Giacomini:2018gxh, Hohn:2018iwn, Vanrietvelde:2018dit, Vanrietvelde:2018pgb, Castro-Ruiz:2019nnl, Hardy:2019cef, Hoehn:2019fsy, Ballesteros:2020lgl, delaHamette:2020dyi, Giacomini:2020ahk, Hoehn:2020epv, Krumm:2020fws, Streiter2020, Tuziemski2020, Yang2020, AliAhmad:2021adn, Castro-Ruiz:2021vnq, Cepollaro:2021ccc, delaHamette:2021iwx, delaHamette:2021oex, delaHamette:2021piz, Giacomini:2021aof, Giacomini:2021gei, Hoehn:2021flk, Mikusch:2021kro, Christodoulou:2022knr, delaHamette:2022cka, Giacomini:2022hco, Kabel:2022cje, Overstreet:2022zgq, Carette:2023wpz, Glowacki:2023nnf, Hoehn:2023axh, Hoehn:2023ehz, Kabel:2023jve, Carrozza:2024smc, DeVuyst:2024fxc, Fewster:2024pur, Kabel:2024lzr} has started to show convergence between the communities. Each of these approaches is best suited to answer a different set of questions. However, in general the QRF concept marks a departure from the usual separation between the observed quantum system and the laboratory which defines a classical reference frame, and aims at developing a relational approach where the reference frames are themselves part of the quantum system under study. 
 
In this work, we focus on the approach introduced in Ref.~\cite{Giacomini:2017zju}. As we are going to review in more details in Section~\ref{sec:QRF}, in this formulation transformations between different QRFs take a similar form as standard transformations between idealised reference frames, promoting the parameter of the transformation to a quantum operator acting on the Hilbert space of the QRF. For example, considering a system of quantum particles and ignoring time evolution, the transformations corresponding to generalised Galilean translations and boosts acting on a particle $B$ are described, respectively, by the following operators~\cite{Giacomini:2017zju, Ballesteros:2020lgl}:
\begin{equation} \label{tybintro}
    \hat{U}_P = e^{\frac{i}{\hbar} \hat{x}_A \otimes \hat{P}_B} \ , \qquad
    \hat{U}_K = e^{\frac{i}{\hbar} \frac{\hat{\pi}_A}{m_A} \otimes \hat{K}_B} \ .
\end{equation}
In this construction, the transformation parameters have a physical interpretation as phase space coordinates $(\hat{x}_A, \hat{\pi}_A)$ of the quantum particle $A$ playing the role of reference frame, while the generators can be represented on the phase space $(\hat{x}_B,\hat{\pi}_B)$ of the quantum particle $B$ which is transformed:~\footnote{In this paper we slightly change notation with respect to Ref.~\cite{Giacomini:2017zju} and call the boost generator $K_B$ instead of $G_B$.}
\begin{equation} \label{pbgb}
    \hat{P}_B = \hat{\pi}_B \ , \qquad
    \hat{K}_B = t \hat{\pi}_B - m_B \hat{x}_B \ .
\end{equation}
This construction generalises the usual Galilean transformations of quantum mechanics, because it allows one to consider, for instance, superpositions of spatial translations and superpositions of Galilean boosts. These generalised Galilean transformations are therefore defined on the tensor product space of two noncommutative algebras acting on different spaces, namely the phase space of particle $A$ and the phase space of particle $B$.

In quantum gravity, it is expected that spacetime acquires quantum properties, such as noncommutativity, when probed at the Planck scale. This can be related to the notion of quantum observers~\footnote{Notice that the notion of observer in quantum information and quantum foundations is not unique. Observers are required to make sense operationally of the outputs of measurements and of probability distributions. Hence, the definition of observer usually requires the specification of a measurement apparatus.} linked by quantum transformations~\cite{Amelino-Camelia:1998jxz, Lizzi:2021bfq}. Mathematically, such transformations can be described within the framework of quantum groups~\cite{Chari:1994pz}, which are a generalisation of Lie groups where the group parameters become noncommutative operators. We recall that quantum groups have also found a wide range of applications to describe the symmetries of, for instance, exactly solvable spin chains, (1+1)-dimensional integrable field theories, nuclear physics and noncommutative geometry (see~\cite{Majid:1988we, Pasquier:1989kd, Gomez:1996az, Bonatsos:1999xj, Szabo:2001kg} and references therein).

Interestingly, quantum groups can be written in a formally similar fashion as the QRF transformations~\eqref{tybintro} by making use of the so-called universal T-matrix formalism~\cite{Fronsdal:1991gf, Fronsdal1994}. In fact, as we discuss in detail in Section~\ref{sec:quantumgroups}, the quantum translations and boosts of a quantum (centrally extended) Galilei group can be written as follows: 
\begin{equation} \label{qgpk}
    \hat{G}_P = e^{\hat{a} \otimes \hat{P}_1} \ , \qquad
    \hat{G}_K = e^{\hat{v} \otimes \hat{K}} \ ,
\end{equation}
where the transformation parameters $\hat{a}$ and $\hat{v}$ belong to some noncommutative algebra. The full algebra of transformation parameters is generated by $(\hat{\theta}, \hat{b}, \hat{a}, \hat{v})$, where $\hat{\theta}$ and $\hat{b}$ are the noncommutative parameters associated respectively to the mass generator $\hat{M}$ and to the time translation generator $\hat{P}_0$. The set of operators $(\hat{M}, \hat{P}_0, \hat{P}_1, \hat{K})$ generate a quantum deformation of the (centrally extended) Galilei Lie algebra, with deformation parameter $\alpha$. This is a crucial difference with respect to the QRF construction, where the generators $(\hat{P}_B, \hat{K}_B)$ belong to the undeformed (centrally extended) Galilei Lie algebra.

Given the striking formal similarity of the two generalisations of reference frame transformations given in~\eqref{tybintro} and~\eqref{qgpk}, it is natural to wonder whether these two notions of reference frames are linked. On the one hand, it is well known how the quantum groups formalism can be used to model the expected `fuzzy' properties of quantum spacetime  (see, for instance,~\cite{Lizzi:2018qaf, Ballesteros:2021bhh, DEsposito:2024wru}), and it provides a richer structure at the formal level compared to the one found so far in the QRFs approach. For example, there exists a large number of nonisomorphic noncommutative algebras $(\hat{\theta}, \hat{b}, \hat{a}, \hat{v})$ fulfilling the quantum group axioms (see~\cite{Opanowicz:1998zz, Ballesteros:1999ew, Opanowicz:1999qp}), and each of these solutions defines a different quantum centrally extended Galilei group. 
However, how quantum groups realise the notion of quantum observers concretely is still largely unexplored.  In particular, and in contrast with the QRF construction, realisations of the noncommutative transformation parameters on the phase space of a (quantum) reference frame  have not been investigated. On the other hand, transformations between QRFs have a direct physical interpretation in terms of the transformations to the relative quantities defined from the perspective of different quantum systems. However, the way in which QRFs may realise symmetry transformations of a quantum spacetime is still mostly unexplored. Our goal here is to provide a more definite physical interpretation to quantum groups and to the quantum deformation parameter $\alpha$, while at the same time contributing to the understanding of the symmetry structures defined by QRF transformations.

Recently, it was shown that, at least in the case of free quantum particles, QRF transformations close a Lie group defined by seven generators \cite{Ballesteros:2020lgl}, which identify a symmetry structure in QRFs. For quantum observers, a mathematically consistent notion of generalised symmetry transformations linking them is provided by quantum groups, which define the symmetry structure of the underlying noncommutative spacetime \cite{Ballesteros:2021inq, Fabiano:2023uhg}. In particular, a natural quantum group to consider in our context is the quantum group deformation of the (centrally extended) Galilei group, i.e.~the quantum Galilei group. The possible realisations of the quantum Galilei group have been fully classified in (1+1) dimensions in Refs.~\cite{Opanowicz:1998zz, Ballesteros:1999ew, Opanowicz:1999qp}.

We here show that QRF transformations, with their associated Lie group as identified in Ref.~\cite{Ballesteros:2020lgl}, can be described in terms of a quantum Galilei group. In particular, we identify a specific realisation of a quantum Galilei group with the symmetry group of QRFs identified in~\cite{Ballesteros:2020lgl}, therefore providing a novel and promising connection between QRF transformations and quantum groups.

The paper is organised as follows. In Section~\ref{sec:Galileigroup}, we review the structure of the Galilei group with a central (mass) extension, which is the symmetry group of free particles in standard quantum mechanics. This sets up the language at the foundation of both the QRF formalism (Section~\ref{sec:QRF}) and the quantum Galilei group construction (Section~\ref{sec:quantumgroups}). In addition, in Section~\ref{sec:quantumgroups} we review the argument showing that the definition of a quantum Galilei group is not unique, because there exists a finite set of possible quantum deformations of the classical Galilei group which give rise to a set of admissible noncommutative algebras with different relations among the noncommutative transformation parameters $(\hat{\theta}, \hat{b}, \hat{a}, \hat{v})$. We argue that, under very mild requirements, there exists a unique quantum Galilei group among all possible realisations, whose noncommutative transformation parameters may correspond to the algebraic properties of the QRF transformation parameters (Subsection~\ref{sec:commutingtime}). Such a  group is the cornerstone of this work, 
and is explicitly defined by the following noncommutative algebra of quantum Galilei group parameters
\begin{equation}
	[\hat{a}, \hat{v}] = \alpha \hat{v} \ , \qquad
	[\hat{a}, \hat{\theta}] = \alpha \hat{\theta} \ , \qquad
	[\hat{v}, \hat{\theta}] = - \frac{1}{2} \alpha \hat{v}^2\ , \qquad
	[\hat{b}, \cdot] = 0 \ .
    \nonumber
\end{equation}
Note that in this algebra the quantum  coordinate $\hat{b}$ commutes with the rest of the generators, and therefore can be considered as a commutative time parameter under any irreducible representation.

Notice that establishing a correspondence between the formal structure of a quantum Galilei group and the physical realisations of QRF transformations is \emph{a priori} not trivial. The crucial ingredient that allows us to find such a link, as discussed in Section~\ref{sec:relation}, is the observation that the group noncommutative transformation parameters$(\hat{\theta}, \hat{b}, \hat{a}, \hat{v})$ can be interpreted as operators on the phase space  of a quantum particle. This leads us to find a representation of these noncommutative parameters on a canonical phase space,  given explicitly by 
\begin{equation}
    \hat{a} = \hat{q} + \gamma \ , \qquad \hat{v} = \phi\, e^{\alpha \hat{p}} \ , \qquad \hat{\theta} = \frac{\phi}{4} (e^{\alpha \hat{p}} \hat{q} + \hat{q} e^{\alpha \hat{p}}) \ , \qquad \hat{b} = \delta \ ,
    \nonumber
\end{equation}
where $(\hat{q}, \hat{p})$ are canonical phase space operators and $\gamma$, $\phi$ and $\delta$ are real parameters, the latter representing the commutative time parameter (Subsection~\ref{sec:QGGphasespace}).
Thanks to this, we uncover the exact correspondence between Galilean QRF transformations and the quantum Galilei group transformations, taken at first order in the quantum deformation parameter $\alpha$, since we then have that $$\hat{a} = \hat{q} + \gamma \ , \quad \hat{v} = \phi (1 + \alpha \hat{p}) + \mathcal{O}(\alpha^2) \ .$$ Therefore, in this limit the commutator  $[\hat{a}, \hat{v}]$ is just a constant, so that we can identify the noncommutative parameters $\hat{a}$ and $\hat{v}$ of the quantum group~\eqref{qgpk} with, respectively, the coordinate $\hat{x}_A$ and velocity $\hat{v}_A \equiv \frac{\hat{\pi}_A}{m_A}$ of the quantum particle defining the QRF transformations~\eqref{tybintro}. As we will show, the derivation of this correspondence requires the quantum deformation parameter $\alpha$ to be proportional to the inverse of the mass $m_A$ of the particle serving as QRF, and its momentum to be renormalised with respect to the momentum in the phase space where quantum transformation parameters are represented (Subsection~\ref{sec:correspondence}).

In Section~\ref{sec:PLgrouplimit} we show that the semiclassical limit of the quantum Galilei group gives rise to a Poisson--Lie Galilei group that can be interpreted as the group of transformations for classical dynamical reference frames, in which the (now commutative) transformation parameters are endowed with a nontrivial Poisson structure. This provides the tools to establish that QRF transformations describe transformations between systems that are in a quantum superposition state of semiclassical states (Section~\ref{sec:QGGasQRF}). We finally conjecture that transformations between QRFs in generic quantum states are described by quantum groups accounting for all orders in the quantum transformation parameter. The concluding Section~\ref{sec:concluding} summarises the most relevant findings here presented, and discusses several open questions.

\sect{The extended Galilei group in (1+1) dimensions}
\label{sec:Galileigroup}

We start by reviewing the basics of the  extended Galilei Lie group, describing the symmetries of a free quantum particle, as formalised in~\cite{Levy1971, Levy-Leblond:1974wdd}. 

The (1+1) centrally extended Galilei algebra $\Gal$ is a four-dimensional real Lie algebra generated by the time translation $\hat{P}_0$, the spatial translation $\hat{P}_1$, the Galilean boost $\hat{K}$, and the operator  representing to the particle's mass  $\hat{M}$. The Lie brackets and Casimir operators of $\Gal$ are given by
\begin{equation} \label{galilei}
	[\hat{P}_0, \hat{P}_1] = 0 \ , \qquad
    [\hat{K}, \hat{P}_0] = \hat{P}_1 \ , \qquad
    [\hat{K}, \hat{P}_1] = \hat{M} \ , \qquad
    [\hat{M}, \cdot] = 0 \ ;
\end{equation}
\begin{equation} \label{bd}
    \hat{\mathcal{C}}_1 = \hat{M} \ , \qquad
    \hat{\mathcal{C}}_2 = \hat{P}_1^2 - 2 \hat{M} \hat{P}_0 \ .
\end{equation}
The (1+1) extended Galilei group transformations are obtained by exponentiating the above-mentioned generators. Therefore, a Galilei group element is
\begin{equation} \label{qgg}
    G = e^{\theta \hat{M}} \, e^{b \hat{P}_0} \, e^{a \hat{P}_1} \, e^{v \hat{K}} \ .
\end{equation}
where the finite transformation parameters $(\theta, b, a, v)$ parametrise the group. The product of two Galilei group elements, corresponding to the composition of two Galilean transformations, must lead to another element of the group: 
\begin{equation} \label{gmult}
    G' \cdot G = e^{\theta' \hat{M}} \, e^{b' \hat{P}_0} \, e^{a' \hat{P}_1} \, e^{v' \hat{K}} 
    \cdot e^{\theta \hat{M}} \, e^{b \hat{P}_0} \, e^{a \hat{P}_1} \, e^{v \hat{K}} 
    = G'' = e^{\theta'' \hat{M}} \, e^{b'' \hat{P}_0} \, e^{a'' \hat{P}_1} \, e^{v'' \hat{K}} \ .
\end{equation}
From this requirement, and by reordering the exponentials on the left-hand side, the following (1+1) Galilei group law is obtained:
\begin{equation} \label{glaw}
    (\theta'', b'', a'', v'') = (\theta' + \theta + v' a + \tfrac{1}{2} v'^2 b, 
    b' + b, a' + a + v' b, v' + v) \ .
\end{equation}
Obviously, the explicit form of the group law depends on the chosen ordering for the exponentials in the group element $\hat{G}$. Our choice corresponds to the one made in~\cite{Levy1971} and leads to the group law provided in Eq.~(3.40) therein. Notice that the physical meaning of the $\theta$ parameter can be understood in terms of the Bargmann phase of standard Galilean quantum mechanics~\cite{Bargmann:1954gh}, obtained from subsequently applying a boost and a translation and their inverses. In fact, using Eq.~\eqref{glaw}, we have 
\begin{equation} \label{bargmannp}
    (0, 0, a, -v) \cdot (0, 0, -a, v) = (v a, 0, 0, 0) \ .
\end{equation}
It is convenient to define the ``physical'' generators $(\hat{M}', \hat{P}_0', \hat{P}_1', \hat{K}')$, which have the correct physical dimensions to describe the symmetries of quantum mechanics. These are related to the ``abstract'' generators $(\hat{M}, \hat{P}_0, \hat{P}_1, \hat{K})$  via $\hat{X}'= - i \hbar \hat{X}$ and satisfy the commutation rules
\begin{equation} \label{galileiphysical}
	[\hat{P}_0', \hat{P}_1'] = 0 \ , \qquad
    [\hat{K}', \hat{P}_0'] = - i \hbar \hat{P}_1' \ , \qquad
    [\hat{K}', \hat{P}_1'] = - i \hbar \hat{M}' \ , \qquad
    [\hat{M}', \cdot] = 0 \ .
\end{equation}
In terms of the physical generators, the Galilei group element~\eqref{qgg} reads
\begin{equation} \label{gih}
    G = e^{\frac{i}{\hbar} \theta \hat{M}'} \, e^{\frac{i}{\hbar} b \hat{P}_0'} \, e^{\frac{i}{\hbar} a \hat{P}_1'} \, e^{\frac{i}{\hbar} v \hat{K}'} \ .
\end{equation}

The physical generators can be represented as operators acting on the noncommutative phase space $[\hat{x}, \hat{\pi}] = i \hbar$ of a quantum-mechanical particle of mass $m$ as follows~\cite{Ballesteros:2020lgl}:
\begin{equation}
	\hat{M}' = m \ , \qquad
	\hat{P}_0' = \frac{\hat{\pi}^2}{2 m} \ , \qquad
    \hat{P}_1' = \hat{\pi} \ , \qquad
    \hat{K}' = t \hat{\pi} - m \hat{x} \ .
\end{equation}
Therefore, these generators have the physical dimensions of mass, energy, momentum, and mass$\times$length, respectively. Accordingly, the group coordinates $(\theta, b, a, v)$ have dimensions
\begin{equation} \label{dimensions}
    [\theta] = L^2 T^{-1} \ , \qquad
    [b] = T \ , \qquad
    [a] = L \ , \qquad
    [v] = L T^{-1} \ .
\end{equation}
These considerations will turn out useful in Section~\ref{sec:relation}, where we provide a physical interpretation of the quantum analogue of the Galilei group.


\sect{Galilean quantum reference frames}
\label{sec:QRF}

In this Section, we review the basics of the formalism for QRFs introduced in~\cite{Giacomini:2017zju} and we recall the result of~\cite{Ballesteros:2020lgl} that QRF transformations in single-particle Galilean quantum mechanics close a Lie group.

In the framework of~\cite{Giacomini:2017zju}, a QRF is associated to a physical system whose coordinates can have quantum properties. The formalism is relational, in the sense that only the relative variables, for instance relative positions, are physically meaningful. To fix the ideas, let us first consider three classical particles: $A$, $B$, and $C$. In this case, the relational nature of the description means that, if one of the three particles, say $C$, identifies the origin of the reference frame, only the relative positions between $C$ and $A$, $x_A^{(C)}$, and between $C$ and $B$, $x_B^{(C)}$, are relevant. Conversely, if the origin is identified with the position of another particle, say $A$, the physically meaningful quantities are the relative positions between $A$ and $C$, $x_C^{(A)}$, and between $A$ and $B$, $x_B^{(A)}$. In this sense, the state of the system and the observables should be understood as properties of the relation between the chosen reference frame (respectively, $C$ or $A$ in the example) and the system under consideration, and not as absolute properties of the system by itself.

Let us now turn to the quantum case and consider $A$, $B$, and $C$ to be quantum particles in single-particle Galilean quantum mechanics. The relational construction we have introduced defines a relational phase space in the perspective of, say, system $C$, i.e.\,$(\hat{x}_A^{(C)}, \hat{\pi}_A^{(C)}, \hat{x}_B^{(C)}, \hat{\pi}_B^{(C)})$. Here, the phase space operators correspond to the position and momenta, respectively, of particles $A$ and $B$ as seen from $C$, and live on an infinite-dimensional Hilbert space $\mathcal{L}(\mathcal{H}_i^{(A)})$, with $i = B, C$, of linear operators acting on the single particle Hilbert space $\mathcal{H}_i^{(A)} \cong L^2 (\mathbb{R})$ on which the vector state of the particle is defined. They satisfy the usual quantum-mechanical commutation relations,
\begin{equation} \label{commkh}
    [\hat{x}_A^{(C)}, \hat{\pi}_A^{(C)}] = i \kappa \ , \qquad
    [\hat{x}_B^{(C)}, \hat{\pi}_B^{(C)}] = i \hbar \ .
\end{equation}
where the two constants $\kappa$ and $\hbar$ have been introduced in order to distinguish explicitly the noncommutativity describing each quantum particle\footnote{In the end, we want to set $\kappa = \hbar$, but keeping them distinct at this level is useful to understand from which quantum system the noncommutative effects arise.}.

The Hamiltonian $\hat{H}_{AB}^{(C)}$ of $A$ and $B$ from the point of view of $C$ is taken to be the free particle Hamiltonian, namely
\begin{equation}
    \hat{H}_{AB}^{(C)} = \frac{(\hat{\pi}_A^{(C)})^2}{2 m_A} + \frac{(\hat{\pi}_B^{(C)})^2}{2 m_B} \ .
\end{equation}
where $m_A$ and $m_B$ are the masses, respectively, of particle $A$ and $B$. Notice that the particles $A$ and $B$ can have any quantum state from the perspective of $C$, and hence also a state with a very large coherent delocalisation in position or momentum. Imagine, for instance, that $A$ has a quantum state as seen from $C$ that is described by a delocalised function in position basis. If this is the case, it is impossible to apply a standard translation on the coordinates of particle $B$ to the reference frame whose origin is at the position of particle $A$, because $A$ is in a quantum superposition of different positions. The formalism for QRFs prescribes that the parameter $a$ of the standard translation acting on the coordinates of system $B$, i.e. $\hat{U}_a = e^{\frac{i}{\hbar}a \hat{P}_B}$, with $\hat{P}_B$ being the generator of the translations, is promoted to a quantum operator on the Hilbert space of the QRF $A$. Physically, this means that the translation amount is controlled by the position of quantum particle $A$, and that translations by different positions can be coherently superposed. An analogous reasoning can be applied to the Galilean boost, when system $A$ is in a delocalised state in momentum basis. Overall, the two QRF transformations corresponding to the superposition of spatial translations and to the superposition of Galilean boosts contain the following generalised translation and boost operators (we drop here the label $(C)$):
\begin{equation} \label{tybb}
    \hat{U}_P = e^{\frac{i}{\hbar} \hat{x}_A \otimes \hat{P}_B} \ , \qquad
    \hat{U}_K = e^{\frac{i}{\hbar} \frac{\hat{\pi}_A}{m_A} \otimes \hat{K}_B} \ .
\end{equation}
where the group parameters $(\hat{x}_A, \hat{\pi}_A)$ are no longer commutative quantities, and $(\hat{P}_B, \hat{K}_B)$ are the translation and boost generators of the Galilei algebra given by
\begin{equation}
    \hat{P}_B = \hat{\pi}_B \ , \qquad
    \hat{K}_B = t \hat{\pi}_B - m_B \hat{x}_B \ .
\end{equation}
It is immediate to notice that, if the parameters of the algebra are quantum operators on the Hilbert space of the QRF $A$, the QRF transformations do not close the usual Galilei group. In~\cite{Ballesteros:2020lgl} it was shown, however, that there is a 7-dimensional Lie group of inertial QRF transformations, and hence that a symmetry structure can be recovered in QRFs. Specifically, the generators of the group are
\begin{equation}\label{QRFoperators}
    \begin{gathered}
        \hat{P}_{AB} = \hat{x}_A \otimes \hat{\pi}_B \ , \qquad 
        \hat{K}_{AB} = \frac{\hat{\pi}_A}{m_A} \otimes \hat{K}_B \ , \\
        \hat{Q}_A = \frac{\hat{\pi}_A^2}{2 m_A} \otimes \mathbb{1}_B \ , \qquad  
        \hat{Q}_B =  \mathbb{1}_A \otimes \frac{\hat{\pi}_B^2}{2 m_B} \ ,\\
        \hat{D}_{A} = \frac{1}{2}(\hat{x}_A \hat{\pi}_A + \hat{\pi}_A \hat{x}_A) \otimes \mathbb{1}_B \ , \qquad 
        \hat{D}_{B} = \mathbb{1}_A \otimes \frac{1}{2}(\hat{x}_B \hat{\pi}_B + \hat{\pi}_B \hat{x}_B) \ , \qquad 
        \hat{T} = \hat{\pi}_A \otimes \hat{\pi}_B \ .
    \end{gathered}
\end{equation}
and they close the following Lie algebra, which in~\cite{Ballesteros:2020lgl} was called the dynamical algebra $\mathcal{D}(7)$:\footnote{Note that $\mathcal{D}(7)$ is a one-parametric family of Lie algebras with parameter $t$. When $t = 0$ the generators $(\hat{P}_{AB}, \hat{K}_{AB}, \hat{D}_A, \hat{D}_B)$ define a Lie subalgebra which is the so-called relational Lie algebra $\mathcal{R}(4)$.}
\begin{equation} \label{eq:dynamicalalgebra}
	\begin{aligned}
		&[\hat{P}_{AB}, \hat{K}_{AB}] = i \hbar \frac{m_B}{m_A} \hat{D}_A - i \kappa \frac{m_B}{m_A} \hat{D}_B + 2 i \kappa \frac{m_B}{m_A} \hat{Q}_B t , 
		&&[\hat{P}_{AB}, \hat{Q}_A] = i \frac{\kappa}{m_A} \hat{T} ,  
		&&[\hat{P}_{AB}, \hat{Q}_B] = 0 , \\
		&[\hat{P}_{AB}, \hat{D}_A] = i \kappa \hat{P}_{AB} , 
		&&[\hat{P}_{AB}, \hat{D}_B] = - i \hbar \hat{P}_{AB} , 
		&&[\hat{P}_{AB}, \hat{T}] = 2 i \kappa m_B \hat{Q}_B , \\
		&[\hat{K}_{AB}, \hat{Q}_A] = 0 , 
		&&[\hat{K}_{AB}, \hat{Q}_B] = - i \frac{\hbar}{m_A} \hat{T} , 
		&&[\hat{K}_{AB}, \hat{D}_A] = - i \kappa \hat{K}_{AB} , \\
		&[\hat{K}_{AB}, \hat{D}_B] = i \hbar \hat{K}_{AB} - 2 i \frac{\hbar}{m_A} \hat{T} t , 
		&&[\hat{K}_{AB}, \hat{T}] = - 2 i \hbar m_B \hat{Q}_A , 
		&&[\hat{Q}_A, \hat{Q}_B] = 0 , \\
		&[\hat{Q}_A, \hat{D}_A] = 2 i \kappa \hat{Q}_A , 
		&&[\hat{Q}_A, \hat{D}_B] = 0 , 
		&&[\hat{Q}_A, \hat{T}] = 0 , \\
		&[\hat{Q}_B, \hat{D}_A] = 0 \ , 
		&&[\hat{Q}_B, \hat{D}_B] = 2 i \hbar \hat{Q}_B , 
		&&[\hat{Q}_B, \hat{T}] = 0 , \\
		&[\hat{D}_A, \hat{D}_B] = 0 , 
		&&[\hat{D}_A, \hat{T}] = i \kappa \hat{T} , 
		&&[\hat{D}_B, \hat{T}] = i \hbar \hat{T} .
	\end{aligned}
\end{equation}
This result confirms that the composition of two QRF transformations is again a QRF transformation, thus solving potential algebraic and conceptual issues.

In this paper, we go a step further, and ask whether the group structure of the QRF transformations can be obtained as the limit of a quantum Galilei group and, if this is the case, whether there is a relation between such a quantum group structure and the 7-dimensional Lie group of~\cite{Ballesteros:2020lgl} generated by the dynamical algebra~\eqref{eq:dynamicalalgebra}. 


\sect{Quantum Galilei groups}
\label{sec:quantumgroups}

In this Section, we review the quantum Galilei groups and discuss their semiclassical limit as  Poisson--Lie groups.

Leaving technicalities to the Appendix, a quantum Galilei group is a fully noncommutative version of the exponential map~\eqref{qgg}, in which the  group coordinates $(\theta, b, a, v)$ are transformed into noncommutative operators $(\hat{\theta}, \hat{b}, \hat{a}, \hat{v})$. As shown in the Appendix, the quantum Galilei group element is defined as the object
\begin{equation} \label{Galpha}
    \hat{G}_\alpha = e^{\hat{\theta} \hat{M}} e^{\hat{b} \hat{P}_0} e^{\hat{a} \hat{P}_1} e^{\hat{v} \hat{K}} \ ,  
\end{equation}
where exponentials are just formal power series and $(\hat{\theta}, \hat{b}, \hat{a}, \hat{v})$ generate a non-Abelian algebra, whose noncommutativity is ruled by a ``quantum deformation parameter'' $\alpha$. Notice that  not all quantum groups can be written in a simple exponential form as in \eqref{Galpha}. In general, the construction of the exponential map for quantum groups may involve $q$-exponential functions~\cite{Fronsdal:1991gf, Fronsdal1994}. Nevertheless, we show in  the Appendix that for the quantum Galilei group we are interested in only ordinary exponentials arise.

This deformation of the algebra of coordinates is such that, when $\alpha \to 0$, the group coordinates become commutative and one recovers the usual extended Galilei group of Section~\ref{sec:Galileigroup}. Importantly, the quantum coordinates $(\hat{\theta}, \hat{b}, \hat{a}, \hat{v})$ commute with the generators $(\hat{M}, \hat{P}_0, \hat{P}_1, \hat{K})$. In order to emphasise this, one can write
\begin{equation} \label{Galpha1}
    \hat{G}_\alpha = e^{\hat{\theta} \otimes \hat{M}} e^{\hat{b} \otimes \hat{P}_0} e^{\hat{a} \otimes \hat{P}_1} e^{\hat{v} \otimes \hat{K}} \ .
\end{equation}
As we mentioned in the introduction, this type of structure strongly resembles the construction of generalised translations and boosts in the QRF approach.

The commutation relations between the quantum group coordinates must satisfy a number of consistency requirements. Clearly, the product of two quantum group elements of the form~\eqref{Galpha1} must be another quantum group element, namely
\begin{equation} \label{gp}
    (e^{\hat{\theta} \otimes \hat{M}}  
    e^{\hat{b} \otimes \hat{P}_0}  
    e^{\hat{a} \otimes \hat{P}_1}  
    e^{\hat{v} \otimes \hat{K}})  
    \cdot  
    (e^{\hat{\theta}' \otimes \hat{M}}  
    e^{\hat{b}' \otimes \hat{P}_0}  
    e^{\hat{a}' \otimes \hat{P}_1}  
    e^{\hat{v}' \otimes \hat{K}})  
    = e^{\hat{\theta}'' \otimes \hat{M}}  
    e^{\hat{b}'' \otimes \hat{P}_0}  
    e^{\hat{a}'' \otimes \hat{P}_1}  
    e^{\hat{v}'' \otimes \hat{K}} \ ,
\end{equation}
where $(\hat{\theta}', \hat{b}', \hat{a}', \hat{v}')$ and $(\hat{\theta}'', \hat{b}'', \hat{a}'', \hat{v}'')$ close the same algebra as $(\hat{\theta}, \hat{b}, \hat{a}, \hat{v})$. Moreover, the composition law of the quantum group coordinates has to be consistent with their commutation rules and must recover the group law of the commutative group coordinates in the limit $\alpha \to 0$, namely it has to be of the form
\begin{equation} \label{glawq}
    (\theta'', b'', a'', v'') = (\theta' + \theta + v' a + \tfrac{1}{2} v'^2 b, b' + b, a' + a + v' b, v' + v) + \mathcal{O}(\alpha) \ .
\end{equation}
Finding the algebra and group law of the quantum group coordinates $(\hat{\theta}, \hat{b}, \hat{a}, \hat{v})$ is nontrivial. Consistency between the two might in fact require that the algebra of the generators $(\hat{M}, \hat{P}_0, \hat{P}_1, \hat{K})$ is also deformed. This means that the algebra \eqref{galilei} acquires additional terms proportional to the same quantum deformation parameter $\alpha$ that governs the noncommutativity of the algebra of the group coordinates. Technically, all the information defining a quantum Galilei group~\eqref{Galpha1} that satisfies all the aforementioned consistency conditions can be obtained by the so-called exponential mapping for quantum groups (or universal T-matrix approach)~\cite{Fronsdal:1991gf, Fronsdal1994}, as we will explicitly show in the Appendix.

In general, several quantum deformations of a given Lie group fulfilling the above consistency conditions can exist. In particular, the Galilei group in (1+1) dimensions admits 26 inequivalent quantum group structures, whose complete classification  was given in~\cite{Opanowicz:1998zz, Ballesteros:1999ew, Opanowicz:1999qp}. By comparing the noncommutative algebras of quantum Galilei group coordinates with the algebra of QRF transformation parameters, we are able to single out a unique quantum Galilei group according to the following two criteria. Given the apparent correspondence of the generalised translation and boost transformations of the QRF approach~\eqref{tybb} with the quantum translation and boost appearing in the quantum group element~\eqref{Galpha}, we require that the quantum transformation parameters $\hat{a}$ and $\hat{v}$ have a nonvanishing commutator, similarly to the QRF transformation parameters $\hat{x}_A$ and $\hat{\pi}_A/m_A$. Moreover, because in the QRF approach time is a commutative parameter, we ask that the time translation parameter $\hat{b}$ of the quantum Galilei group~\eqref{Galpha} is a central element of the algebra of quantum group coordinates (i.e. it commutes with all other elements of the algebra). 

The unique structure of the quantum Galilei group singled out by the two requirements above is discussed in Subsection~\ref{sec:commutingtime}. Since the classification of quantum Galilei groups is based on the one-to-one correspondence between quantum deformations of a given Lie group and their Poisson--Lie structures (which can be considered as the semiclassical analogues of quantum groups~\cite{Chari:1994pz}) we first introduce in the following Subsection the Poisson--Lie Galilei group underlying the quantum Galilei group with which we will work.


\subsection{The semiclassical analogue: Poisson--Lie Galilei groups}
\label{sec:PLgroup}

To construct a quantum Galilei group, we first need to introduce its semiclassical counterpart, the Poisson--Lie group. The group parameters  $(\theta, b, a, v)$ of a Poisson--Lie Galilei group,
\begin{equation} \label{qgg2}
    G_\alpha = e^{\theta \hat{M}} e^{b \hat{P}_0} e^{a \hat{P}_1} e^{v \hat{K}} \ ,
\end{equation}
are classical coordinate functions, and hence commutative, but they display dynamical properties, in the sense that they define the fundamental brackets of a Poisson algebra structure which is again ruled by the parameter $\alpha$. This Poisson--Lie structure on the Galilei group allows for the introduction of Hamiltonian dynamics on the space of smooth functions on the group parameters, specified by some Hamiltonian $H(\theta, b, a, v)$ and defining evolution with respect to a parameter $s$:
\begin{equation}
    \dot{x} = \frac{dx}{ds} = \{x, H\} \ , \qquad x = \{\theta, b, a, v\} \ .  
\end{equation}
Similarly to what we discussed for the quantum Galilei group, there are analogous compatibility conditions between the Poisson structure on the group coordinates and the Galilei group law~\eqref{gmult}. Namely, the Poisson brackets for the $(\theta'', b'', a'', v'')$ dynamical variables must be formally the same as the Poisson brackets for $(\theta, b, a, v)$ and for $(\theta', b', a', v')$, once they are computed from \eqref{glaw} taking into account that\footnote{This follows from observing that $(x, y)$ and $(x', y')$ live on different copies of the group and therefore the Poisson bracket between them vanishes.} 
\begin{equation} \label{bb}
    \{ x' x, y' y \} = \{ x', y' \} \, x \, y + x' \, y' \, \{ x, y \} \ .  
\end{equation}

The construction of a compatible Poisson--Lie structure on a given group is in general not unique. A classification of the 26 nonequivalent Poisson--Lie structures in the centrally extended Galilei group of (1 + 1) dimensions is provided in~\cite{Opanowicz:1998zz} (see also~\cite{Ballesteros:1999ew} for the classification of the corresponding Lie bialgebra structures of the centrally extended Galilei Lie algebra, which are in one-to-one correspondence with Poisson--Lie structures~\cite{Drinfeld:1983ky}). As mentioned previously, we are able to single out one among these structures by asking compatibility with the QRFs defined in Section~\ref{sec:QRF}. The criteria for the commutators of the quantum group coordinates that we exposed above translate into similar requirements on the dynamical parameters of the associated Poisson--Lie structure. In particular, we ask that the bracket $\{a, v\}$ is nonvanishing and that the time coordinate $b$ Poisson-commutes with the other coordinates. The reasons for these two requirements are that, in the undeformed case of QRF transformations, $a$ and $v$ are associated, respectively, to the position and momentum of the QRF. The element $b$ plays the role of time, which is not modified in Galilean QRFs. Hence, it is a natural choice to keep it as a commutative element of the algebra. These conditions uniquely identify one Poisson--Lie structure among the ones provided in~\cite{Opanowicz:1998zz}, which is given by 
\begin{equation} \label{plbracket1}
    \{a, v\} = \alpha v \ , \qquad \{a, \theta\} = \alpha \theta + \beta_1 \alpha^2 v \ , \qquad \{v, \theta\} = - \frac{1}{2} \alpha v^2 \ , \qquad \{b, \cdot\} = 0 \ ,
\end{equation}
where $\alpha$ and $\beta_1$ are real free parameters that govern the Poisson-noncommutativity. While $\alpha$ is essential in defining a compatible Poisson--Lie structure, the parameter $\beta_1$ can be set to zero without spoiling the construction. We notice that in the classification of Poisson--Lie Galilei groups~\cite{Opanowicz:1998zz} there is no Poisson--Lie structure in which $\{a, v\}$ is proportional to a central element (a constant), as it would be if we identified $a$ and $v$ with some phase space coordinates $q$ and $p$. This excludes the possibility of finding a straightforward quantum group counterpart of the QRF construction. Nevertheless, we will show in the following that a connection between $\{a, v\}$ and $\{q, p\}$ does exist. This will provide a neat physical link between QRF parameters and quantum group transformations.

We take the algebra~\eqref{plbracket1} as our starting point in the definition of a quantum Galilei group,  based on the property that there exists a one-to-one correspondence between Poisson--Lie structures on a group and its quantum deformations.


\subsection{A quantum Galilei group with commuting time}
\label{sec:commutingtime}

In this Subsection, we construct the quantum counterpart $\hat{G}_\alpha$ of the Poisson--Lie Galilei group $G_\alpha$ with Poisson--Lie structure~\eqref{plbracket1} by means of the exponential map for quantum groups, also called Hopf algebra dual form or universal T-matrix, see~\cite{Fronsdal:1991gf, Fronsdal1994, Bonechi:1993sn, Ballesteros1995} and references therein.

As described in detail in the Appendix, we start from the (Hopf) algebra of ordered monomials of quantum Galilei coordinates   $(\hat{\theta}, \hat{b}, \hat{a}, \hat{v})$ generating the algebra~\eqref{commrules} coming from the quantisation of the Poisson--Lie structure~\eqref{plbracket1} (with $\beta_1 = 0$).  These determine the ordering of the quantum group element~\eqref{Galpha1}. Thus, we choose as basis elements
\begin{equation}
    \hat{z}_{abcd} = \hat{\theta}^a \hat{b}^b \hat{a}^c \hat{v}^d \ ,  
\end{equation}
with the exponents $a, b, c, d$ being nonnegative integers.

The universal T-matrix is  defined via the Hopf algebra dual form,
\begin{equation} \label{hadfa}
    T = \sum_{abcd} \hat{z}_{abcd} \otimes \hat{Z}^{abcd} \ ,
\end{equation}
where $\hat{Z}^{abcd}=\frac{\hat{M}^a}{a!} \frac{\hat{P}_0^b}{b!} \frac{\hat{P}_1^c}{c!} \frac{\hat{K}^d}{d!}$ is  the generic element of the dual basis, such that
\begin{equation} \label{dualgeneratorsa}
    \hat{Z}^{1000} = \hat{M} \ , \qquad \hat{Z}^{0100} = \hat{P}_0 \ , \qquad \hat{Z}^{0010} = \hat{P}_1 \ , \qquad \hat{Z}^{0001} = \hat{K} \ ,
\end{equation}
are the generators of the quantum Galilei algebra, see the Appendix. From this result, the Galilei universal T-matrix~\eqref{hadfa} is straightforwardly proven to be given as an ordered product of ordinary exponentials,
\begin{equation} \label{tgal}
    T = e^{\hat{\theta} \otimes \hat{M}}  
    e^{\hat{b} \otimes \hat{P}_0}  
    e^{\hat{a} \otimes \hat{P}_1}  
    e^{\hat{v} \otimes \hat{K}}  
    \equiv \hat{G}_\alpha \ ,
\end{equation}
where the quantum Galilei generators $(\hat{M}, \hat{P}_0,\hat{P}_1, \hat{K})$ close an $\alpha$-deformed algebra. Indeed, in the limit $\alpha \to 0$ the universal T-matrix~\eqref{tgal} leads to the exponential map for the usual centrally extended Galilei group~\eqref{qgg}, where the group parameters are commutative and the undeformed Galilei Lie algebra is recovered.

As we mentioned, the quantum group coordinates appearing in \eqref{tgal} satisfy commutators induced by the quantisation of the Poisson--Lie structure~\eqref{plbracket1} (with $\beta_1 = 0$). This quantisation procedure is analogous to the quantisation of the usual Poisson brackets between the phase space dynamical variables, $\{q, p\} = 1$, which turns them into a commutator $[\hat{q}, \hat{p}] = i \hbar$. However, the quantisation parameter used in this context is in principle a different constant than $\hbar$, which we call $\kappa$ (see also discussions in~\cite{Ballesteros:2021inq, Lizzi:2021bfq, Gubitosi:2021itz}). Therefore, the quantisation of~\eqref{plbracket1} leads to
\begin{equation} \label{qga}
    [\hat{a}, \hat{v}] = i \kappa \alpha \hat{v} \ , \qquad [\hat{a}, \hat{\theta}] = i \kappa \alpha \hat{\theta} \ , \qquad [\hat{v}, \hat{\theta}] = - \frac{i}{2} \kappa \alpha \hat{v}^2 \ , \qquad [\hat{b}, \cdot] = 0 \ .
\end{equation}
The physical interpretation of both the deformation parameter $\alpha$ and the quantisation parameter $\kappa$ will become apparent in the following Sections. We stress that the commutators~\eqref{qga} are compatible with the undeformed group law~\eqref{gp}-\eqref{glawq}, with no further corrections in terms of $\alpha$:\footnote{The group law can be formally defined in terms of a ``coproduct'' of the quantum group coordinates. See the Appendix for details.}
\begin{equation} \label{gl}
    (\theta'', b'', a'', v'') = (\theta' + \theta + v' a + \tfrac{1}{2} v'^2 b, 
    b' + b, a' + a + v' b, v' + v) \ .
\end{equation}

On the other hand, the Hopf algebra duality between noncommutative coordinates and quantum algebra generators given by~\eqref{dualgeneratorsa} implies that the algebra of generators of the quantum Galilei group has to be $\alpha$-deformed in order to be compatible with~\eqref{qga}. Such deformation is unique and leads to\footnote{Also for the generators, there is an associated deformed coproduct, which is discussed in the Appendix.}
\begin{equation} \label{qalg}
    [\hat{P}_0, \hat{P}_1] = 0 \ , \qquad [\hat{K}, \hat{M}] = \frac{i}{2} \kappa \alpha e^{- i \kappa \alpha \hat{P}_1} \hat{M}^2 \ ,  
    \qquad [\hat{K}, \hat{P}_0] = \frac{1 - e^{- i \kappa \alpha \hat{P}_1}}{i \kappa \alpha} \ , \qquad [\hat{K}, \hat{P}_1] = e^{- i \kappa \alpha \hat{P}_1} \hat{M} \ ,
\end{equation}
with the Casimir operators deformed accordingly:
\begin{equation} \label{cu}
    \hat{\mathcal{C}}_1 = \left(\frac{\sinh(\frac{i}{4} \kappa \alpha \hat{P}_1)}{\frac{i}{4} \kappa \alpha}\right)^2 - 2 e^{- \frac{i}{2} \kappa \alpha \hat{P}_1} \hat{M} \hat{P}_0 \ , \qquad  
    \hat{\mathcal{C}}_2 = \hat{M} e^{- \frac{i}{2} \kappa \alpha \hat{P}_1} \ .
\end{equation}
Note that, as it should be, the limit $\alpha \to 0$ leads to the undeformed Galilei Lie algebra and its usual Casimir operators.\footnote{One might notice that formally also the limit $\kappa \to 0$ seems to produce the same result. However, as we discuss in detail in the following Sections, while $\alpha$ is the deformation parameter, which induces the deformation of the algebra of quantum group coordinates and generators, $\kappa$ is a quantisation parameter, and its vanishing limit corresponds to the Poisson--Lie group limit studied in~\ref{sec:PLgrouplimit}. The different roles of the two parameters $\alpha$ and $\kappa$ become more apparent within the phase space realisation discussed in the following Section.} 

Summarising, the quantum Galilei group with commuting time we are dealing with is defined by the noncommutative algebra~\eqref{qga} of group parameters and the deformed Galilei algebra of generators~\eqref{qalg}. Both of them are fully compatible through the Hopf algebra duality underlying the definition of the universal T-matrix~\eqref{tgal}, which can be thought of as the complete presentation of a quantum Galilei group element $\hat{G}_\alpha$.

Before proceeding to discuss the relation of this quantum Galilei group with the Galilean QRF transformations, we would like to briefly comment on the role of quantum groups in describing physical systems.

Interestingly, the quantum Galilei algebra~\eqref{qalg} was firstly introduced in~\cite{Bonechi:1992cb, Bonechi:1992ye}, in a different basis and making use of a different representation, as the symmetry describing magnon excitations in (1+1)-dimensional Heisenberg spin chains. In that case, the quantum deformation parameter $\alpha$ is related to the lattice spacing of the system. In the context under study here, we will show that the quantum deformation parameter can be related to the mass of the quantum particle playing the role of a quantum reference frame. It should not be surprising that the same quantum group can be relevant in different physical frameworks. It simply means that those frameworks have the same symmetry properties. What differs is the realisation of the group parameters and generators on the variables of the system under consideration, and the relation of the quantum deformation parameter to the relevant physical quantities.
As another example, in a quantum gravity setting, quantum Galilei coordinates $(\hat{b}, \hat{a})$ are interpreted, respectively, as time and space coordinates $(\hat{x}^0, \hat{x}^1)$ of a quantum Galilean spacetime. Therefore, if we consider the algebra ~\eqref{qga} in this kinematical context, the commutator of time and space coordinates vanishes, thus giving rise to a commutative Galilean spacetime with quantum Galilean symmetries. Nevertheless, this is no longer the case for other quantum Galilei groups, like for instance the so-called $\kappa$-Galilei quantum group, which has been used to describe the symmetries of a quantum Galilean spacetime  with noncommuting time of the form (see~\cite{Ballesteros:2020uxp} and references therein)
$$
[\hat{x}^0, \hat{x}^1] = - \ell \, \hat{x}^1 \ ,
$$ where in this framework the quantum deformation parameter is $\alpha \equiv \ell$, and $\ell$ is identified with the Planck length. We recall that this spacetime can be obtained as the nonrelativistic limit of the $\kappa$-Minkowski noncommutative spacetime, which has been been extensively considered in the quantum gravity literature~\cite{Lizzi:2018qaf}.


\sect{Relation between the Galilean QRF  and the quantum Galilei group transformations}
\label{sec:relation}

Having reviewed the main ingredients of our analysis, we can now characterise the relation between Galilean QRFs and the quantum Galilei group. Because QRF transformations are realised on the phase space of quantum systems, the first step of our construction consists in defining a phase space representation for quantum Galilei transformations (to the best of our knowledge, phase space realisations of universal T-matrix representations of quantum groups have not been considered so far in the literature). This allows us to directly compare them to the QRF formalism and at the same time to shed light on the physical interpretation of quantum group transformations. Thanks to the phase space representation, we can also compare the action of quantum Galilei transformations and Galilean QRF transformations on the phase space of free particles. We find a correspondence between the two kinds of transformations when the quantum group transformations are considered at first order in the quantum deformation parameter $\alpha$, and upon identifying this parameter with the inverse of the mass of the particle playing the role of QRF.


\subsection{Phase-space realisation of the quantum Galilei group}
\label{sec:QGGphasespace}

As done in Section~\ref{sec:Galileigroup} for the classical Galilei group, we transform the abstract quantum Galilei generators of the previous Section into ``physical'' generators by defining $X' = - i \hbar X$. In this way, we get from~\eqref{Galpha1} the analogue of~\eqref{gih} in the quantum group setting:
\begin{equation} \label{quantumG}
    \hat{G}_\alpha = e^{\frac{i}{\hbar} \hat{\theta} \otimes \hat{M}'}
    e^{\frac{i}{\hbar} \hat{b} \otimes \hat{P}_0'}
    e^{\frac{i}{\hbar} \hat{a} \otimes \hat{P}_1'}
    e^{\frac{i}{\hbar} \hat{v} \otimes \hat{K}'} \ ,
\end{equation}
where the noncommutative coordinates fulfil~\eqref{qga} and the quantum algebra of the physical generators is obtained from~\eqref{qalg},
\begin{equation} \label{qalgqrfkappa}
    [\hat{P}_0', \hat{P}_1'] = 0 \ , \, [\hat{K}', \hat{M}'] = \frac{i}{2} \kappa \alpha e^{\frac{\kappa}{\hbar} \alpha \hat{P}_1'} (\hat{M}')^2 \ , \, [\hat{K}', \hat{P}_0'] = i \hbar \frac{1 - e^{\frac{\kappa}{\hbar} \alpha \hat{P}_1'}}{\frac{\kappa}{\hbar} \alpha} \ , \, [\hat{K}', \hat{P}_1'] = - i \hbar e^{\frac{\kappa}{\hbar} \alpha \hat{P}_1'} \hat{M}' \ .
\end{equation}
We call this the physical algebra of quantum generators because in the limit  $\alpha \to 0$ it reduces to the physical algebra of standard Galilei transformations~\eqref{galileiphysical}. For what follows, it is important to emphasise the distinction between the quantisation of the group coordinates~\eqref{qga}, ruled by the constant $\kappa$ defined in the previous Section, and the quantisation of the phase space of the (quantum) Galilei generators, ruled by the Planck constant $\hbar$. Moreover, $\alpha$ is the parameter governing both the deformation of the algebra of the quantum group coordinates and of the group generators. As can be easily checked by dimensional analysis of the transformation parameters \eqref{dimensions}, $\alpha$ has the dimension of an inverse momentum, which is also consistent with \eqref{qalgqrfkappa}.

In order to make a suitable connection with the QRF approach, we represent the algebra of generators~\eqref{qalgqrfkappa} on the phase space $[\hat{q}_B, \hat{p}_B] = i \hbar$, and the algebra of quantum group coordinates~\eqref{qga} on the phase space $[\hat{q}_A, \hat{p}_A] = i \kappa$. Notice that we assume different quantisation constants for the two phase spaces, in accordance to the fact that in the QRF construction of Section~\ref{sec:QRF} we used two different constants to characterise the phase spaces of the quantum particles playing the role of either the reference frame or the transformed system, see Eq.~\eqref{commkh}.

A suitable phase space realisation of the coordinates reads 
\begin{equation} \label{psncc}
    \hat{a} = \hat{q}_A + \gamma \ , \qquad \hat{v} = \phi e^{\alpha \hat{p}_A} \ , \qquad \hat{\theta} = \frac{\phi}{4} (e^{\alpha \hat{p}_A} \hat{q}_A + \hat{q}_A e^{\alpha \hat{p}_A}) \ , \qquad \hat{b} = \delta \ .
\end{equation}
This realisation does not impose any constraint on $\alpha$, and similarly $\phi$ is a free parameter with the same dimensions as $\hat{v}$, that is, a velocity. The other free parameters $(\gamma, \delta)$ label the realisation, since the Casimir operators of the algebra of quantum coordinates,
\begin{equation} \label{casimirsncc}
    I_1 = \hat{a} - (\hat{v}^{-1} \hat{\theta} + \hat{\theta} \hat{v}^{-1}) \ , \qquad I_2 = \hat{b} \ ,
\end{equation}
have eigenvalues $(\gamma, \delta)$ on a given representation~\eqref{psncc}, which implies that this is a generic realisation of the algebra~\eqref{qga} up to canonical transformations in the quantum phase space $[\hat{q}_A, \hat{p}_A] = i \kappa$. These parameters have dimensions of length and time, respectively. Note that because of the noncommutative nature of the transformation parameters and of the phase space coordinates, the given ordering both in the phase space representation and in the first Casimir operator is essential.

The representation of the quantum algebra of generators~\eqref{qalgqrfkappa} on the phase space  $[\hat{q}_B, \hat{p}_B] = i \hbar$ reads
\begin{equation} \label{new_eq}
    \begin{aligned}
        \hat{M}' &= m_B e^{-\frac{\kappa}{2 \hbar} \alpha \hat{p}_B} \ ,  
        &\qquad \hat{P}_0' &= \frac{1}{m_B (\frac{\kappa}{2 \hbar} \alpha)^2} \left(\cosh\left(\frac{\kappa}{2 \hbar} \alpha \hat{p}_B \right) - 1\right) + \xi \ , \\  
        \hat{P}_1' &= \hat{p}_B \ ,  &\qquad \hat{K}' &= - \frac{m_B}{2}\left(e^{\frac{\kappa}{2 \hbar} \alpha \hat{p}_B} \hat{q}_B + \hat{q}_B e^{\frac{\kappa}{2 \hbar} \alpha \hat{p}_B}\right) + t \hat{p}_B \ ,  
    \end{aligned}
\end{equation}
where $\xi$ is an arbitrary constant with the dimensions of an energy. By substituting this realisation into the deformed Casimirs~\eqref{cu}, one finds 
\begin{equation} \label{cu2}
    \hbar^2 \hat{\mathcal{C}}_1 = 2 e^{\frac{\kappa}{2 \hbar} \alpha \hat{P}'_1} \hat{M}' \hat{P}_0'- \left(\frac{\sinh(\frac{\kappa}{4 \hbar} \alpha \hat{P}_1')}{\frac{\kappa}{4 \hbar} \alpha} \right)^2 = 2 m_B \xi \ , \qquad \hbar \hat{\mathcal{C}}_2 = i \hat{M}' e^{\frac{\kappa}{2 \hbar} \alpha \hat{P}_1'} = i m_B \ .
\end{equation}
From this, comparing to the standard Galilean Casimir $\hat{\mathcal{C}}_2 \equiv 2 \hat{M}' \hat{P}_0' - \hat{P}_1'^2 = 2 m u$, one can see that the physical interpretation of $\xi$ is that of the internal energy of the particle, which can be set to zero (this corresponds, in the standard case, to identifying $\hat{P}_0'$ with the kinetic energy).

Putting everything together, the quantum group element~\eqref{quantumG} can be written in terms of the two sets of phase space coordinates $(\hat{q}_A, \hat{p}_A)$ and $(\hat{q}_B, \hat{p}_B)$. To prepare ourselves to find a connection with the QRF transformations~\eqref{tybb}-\eqref{QRFoperators}, we notice that the operators in the exponentials of~\eqref{quantumG} can be written as follows: 
\begin{equation} \label{expT}
    \begin{aligned}
        \hat{M}_\alpha^{AB} &\equiv \hat{\theta} \otimes \hat{M}' = \frac{\phi}{4} (e^{\alpha \hat{p}_A} \hat{q}_A + \hat{q}_A e^{\alpha \hat{p}_A}) \otimes m_B e^{- \frac{\kappa}{2 \hbar} \alpha \hat{p}_B} \ , \\
        \hat{P}_{0, \alpha}^{AB} &\equiv \hat{b} \otimes \hat{P}_0' = \delta \otimes \left[\frac{1}{m_B (\frac{\kappa}{2 \hbar} \alpha)^2} \left(\cosh\left(\frac{\kappa}{2 \hbar} \alpha \hat{p}_B\right) - 1\right) + \xi\right] \ , \\
        \hat{P}_{1, \alpha}^{AB} &\equiv \hat{a} \otimes \hat{P}_1' = (\hat{q}_A + \gamma) \otimes \hat{p}_B \ , \\
        \hat{K}_\alpha^{AB} &\equiv \hat{v} \otimes \hat{K}' = - \phi e^{\alpha \hat{p}_A} \otimes \left(\frac{m_B}{2} (e^{\frac{\kappa}{2 \hbar} \alpha  \hat{p}_B} \hat{q}_B + \hat{q}_B e^{\frac{\kappa}{2 \hbar} \alpha \hat{p}_B}) - t \hat{p}_B\right) \ .
    \end{aligned}
\end{equation}
This allows us to define the following operators that, as we will show in the following, can be directly related to the QRF operators defined in~\cite{Ballesteros:2020lgl}:
\begin{equation} \label{eq:Qgroupele}
    \begin{aligned}
        \hat{U}_M^{\alpha} &\equiv e^{\frac{i}{\hbar} \hat{M}_\alpha^{AB}} = e^{\frac{i}{\hbar} \hat{\theta} \otimes \hat{M}'} \ , &\qquad \hat{U}_{P_0}^\alpha &\equiv e^{\frac{i}{\hbar} \hat{P}_{0, \alpha}^{AB}} = e^{\frac{i}{\hbar} \hat{b} \otimes \hat{P}_0'} \ , \\
        \hat{U}_{P_1}^\alpha &\equiv e^{\frac{i}{\hbar} \hat{P}_{1, \alpha}^{AB}} = e^{\frac{i}{\hbar} \hat{a} \otimes \hat{P}_1'} \ , &\quad \hat{U}_K^\alpha &\equiv e^{\frac{i}{\hbar} \hat{K}_\alpha^{AB}} = e^{\frac{i}{\hbar} \hat{v} \otimes \hat{K}'} \ .
    \end{aligned}
\end{equation}

We mentioned in Section~\ref{sec:Galileigroup} that in standard Galilean quantum mechanics the $\theta$ parameter associated to the generator $M$ can be interpreted in terms of the Bargmann phase~\cite{Bargmann:1954gh}. As shown in~\eqref{bargmannp}, the $\theta$ parameter is just the final result of the transformation that comes out from subsequently applying a boost $v$ and a translation $a$ and their inverses, and its value is $\theta = a v$. In the quantum Galilei group case, the representation of $\hat{a}, \hat{v}$ and $\hat{\theta}$ in~\eqref{psncc} (for $\gamma=0$) shows that such identification between the operators $\hat{\theta}$ and the product $\hat{a} \hat{v}$ is preserved (up to symmetrisation). Therefore, $\hat{\theta}$ can be interpreted as the ``quantum version'' of the Bargmann phase, and as we will see in the following, the operator $e^{\frac{i}{\hbar} \hat{\theta} \otimes \hat{M}'}$ will be essential in order to establish a direct relationship between the quantum Galilei group and the QRF transformations.


\subsection{Connection between the quantum Galilei group and the group of QRF transformations} \label{sec:correspondence}

The two sets of phase space coordinates $(\hat{q}_A, \hat{p}_A)$ and $(\hat{q}_B, \hat{p}_B)$ on which we have represented the quantum Galilei group can be interpreted as the coordinates of two quantum particles, so that the resemblance to the QRF formalism is not just at a formal level. In this Subsection, we show that this identification is especially meaningful when considering only the first order in the quantum deformation parameter $\alpha$. 

Within this approximation, the representation of the quantum group coordinates reads
\begin{equation} \label{parametersrealization}
    \hat{a} = \hat{q}_A + \gamma \ , \quad \hat{v} = \phi (1 + \alpha \hat{p}_A) + \mathcal{O}(\alpha^2) \ , \quad \hat{\theta} = \frac{\phi}{2} (\hat{q}_A + \frac{\alpha}{2}(\hat{p}_A \hat{q}_A + \hat{q}_A \hat{p}_A) ) + \mathcal{O}(\alpha^2) \ , \quad \hat{b} = \delta \ .
\end{equation}
Recognising that $\hat{v}$ has the physical dimension of a velocity, we can define a ``renormalised'' momentum variable\footnote{In the second step we have approximated the expression of $\hat{p}_A'$ to the first order in $\alpha$.} 
\begin{equation}
    \hat{p}'_A = m_A \hat{v} \simeq m_A \phi (1 + \alpha \hat{p}_A) \ .
\end{equation}
Furthermore, taking the Ansatz
\begin{equation} \label{eq:ansatz}
    \alpha = \frac{1}{m_A \phi} \ ,
\end{equation}
we obtain 
\begin{equation} \label{eq:paprime}
    \hat{p}_A' \simeq m_A \phi + \hat{p}_A \ .
\end{equation}
Hence, we notice that $\phi$ (the parameter with dimensions of a velocity which we introduced when representing the quantum Galilei group transformation parameters, Eq.~\eqref{psncc}) rescales  the momentum $\hat{p}_A$ to the renormalised momentum $\hat{p}'_A$. Alternatively, we can express these quantities as velocities by defining  $\hat{v}_A \equiv \frac{\hat{p}_A}{m_A}$, so that 
\begin{equation}
    \hat{v} = \phi + \hat{v}_A \ .
\end{equation}
This renormalised momentum variable brings us much closer to identifying the quantum Galilei group elements with the QRF group elements, as we show in the following. 

Again, working at first order in $\alpha$, the representation of the quantum Galilei generators reads
\begin{equation} \label{generatorsrepresentation}
    \begin{aligned}
        \hat{M}' &= m_B \left(1 - \frac{\kappa}{2 \hbar} \alpha \hat{p}_B\right) + \mathcal{O}(\alpha^2) \ ,  
        &\qquad \hat{P}_0' &= \frac{\hat{p}_B^2}{2 m_B} + \xi + \mathcal{O}(\alpha^2) \ , \\
        \hat{P}_1' &= \hat{p}_B \ ,  &\qquad \hat{K}' &= - m_B \left(\hat{q}_B + \frac{\kappa}{4 \hbar} \alpha (\hat{p}_B \hat{q}_B + \hat{q}_B \hat{p}_B) \right) + t \hat{p}_B + \mathcal{O}(\alpha^2) \ ,  
    \end{aligned}
\end{equation}

Using these expressions, the exponents~\eqref{expT} of the quantum Galilei group~\eqref{quantumG} read, again at first order in $\alpha$,
\begin{equation} \label{linearpsr}
    \begin{aligned}
        \hat{M}_\alpha^{AB} &= \frac{\phi}{2} \hat{q}_A \otimes m_B \left(1 - \frac{\kappa}{2 \hbar} \alpha \hat{p}_B\right) + \frac{\phi}{4} \alpha (\hat{p}_A \hat{q}_A + \hat{q}_A \hat{p}_A) \otimes m_B \ , \\
        \hat{P}_{0, \alpha}^{AB} &= \delta \otimes \left(\frac{\hat{p}_B^2}{2 m_B} + \xi\right) \ , \\
        \hat{P}_{1, \alpha}^{AB} &= (\hat{q}_A + \gamma) \otimes \hat{p}_B \ , \\
        \hat{K}_\alpha^{AB} &= - \phi (1 + \alpha \hat{p}_A) \otimes (m_B \hat{q}_B - t \hat{p}_B) - \phi \otimes \frac{\kappa}{4 \hbar} \alpha m_B (\hat{p}_B \hat{q}_B + \hat{q}_B \hat{p}_B) \ .
    \end{aligned}
\end{equation}
When writing these in terms of the renormalised momentum~\eqref{eq:paprime} we find expressions that are very close to the ones obtained in the QRF approach: 
\begin{equation} \label{qgelementsansatz}
    \begin{aligned}
    \hat{M}_{\alpha}^{AB} &= - \frac{1}{4} \frac{\kappa}{\hbar} \frac{m_B}{m_A} (\hat{q}_A \otimes \hat{p}_B) + \frac{1}{4} \frac{m_B}{m_A} (\hat{p}_A' \hat{q}_A + \hat{q}_A \hat{p}_A') \otimes \mathbb{1}_B \ , \\
    \hat{P}_{0, \alpha}^{AB} &= \delta \mathbb{1}_A \otimes \left(\frac{\hat{p}_B^2}{2 m_B} + \xi \mathbb{1}_B\right) \ , \\
    \hat{P}_{1, \alpha}^{AB} &= (\hat{q}_A + \gamma \mathbb{1}_A) \otimes \hat{p}_B \ , \\
    \hat{K}_{\alpha}^{AB} &= 
    \frac{\hat{p}_A'}{m_A} \otimes (t \hat{p}_B - m_B \hat{q}_B) - \frac{1}{4} \frac{\kappa}{\hbar} \frac{m_B}{m_A} \mathbb{1}_A \otimes (\hat{p}_B \hat{q}_B + \hat{q}_B \hat{p}_B) \ .
    \end{aligned}
\end{equation}
Indeed, by comparing the four operators~\eqref{qgelementsansatz} with the phase space realisation of the QRF operators defining the dynamical Lie algebra $\mathcal{D}(7)$~\eqref{QRFoperators}, which we rewrite here for convenience, 
\begin{equation}
    \begin{gathered}
        \hat{P}_{AB} = \hat{x}_A \otimes \hat{\pi}_B \ , \qquad 
        \hat{K}_{AB} = \frac{\hat{\pi}_A}{m_A} \otimes (t \hat{\pi}_B - m_B \hat{x}_B) \ , \\
        \hat{Q}_A = \frac{\hat{\pi}_A^2}{2 m_A} \otimes \mathbb{1}_B \ , \qquad  
        \hat{Q}_B =  \mathbb{1}_A \otimes \frac{\hat{\pi}_B^2}{2 m_B} \ ,\\
        \hat{D}_{A} = \frac{1}{2}(\hat{x}_A \hat{\pi}_A + \hat{\pi}_A \hat{x}_A) \otimes \mathbb{1}_B \ , \qquad 
        \hat{D}_{B} = \mathbb{1}_A \otimes \frac{1}{2}(\hat{x}_B \hat{\pi}_B + \hat{\pi}_B \hat{x}_B) \ , \qquad 
        \hat{T} = \hat{\pi}_A \otimes \hat{\pi}_B \ ,
    \end{gathered}
\end{equation}
one finds the following correspondences upon identifying the QRF phase space variables $(\hat{x}_A, \hat{\pi}_A, \hat{x}_B, \hat{\pi}_B)$ with the quantum Galilei group phase space coordinates $(\hat{q}_A, \hat{p}_A', \hat{q}_B, \hat{p}_B)$:
\begin{itemize}
\item 
$\hat{M}_\alpha^{AB}$ is a linear combination of $\hat{D}_A$ and  $\hat{P}_{AB}$: 
\begin{equation} \label{eq:M_alpha_AB}
	\hat{M}_\alpha^{AB} = - \frac{1}{4} \frac{\kappa}{\hbar} \frac{m_B}{m_A} \hat{P}_{AB} + \frac{1}{2} \frac{m_B}{m_A} \hat{D}_A \ .
\end{equation}

\item 
$\hat{P}_{0, \alpha}^{AB}$ is a linear combination of the time translation generator $\hat{Q}_B$ and the identity operator related to the internal energy of the system, see discussion after~\eqref{cu2}. This last term might be set to zero by setting the internal energy $\xi = 0$ to make the two time translation generators proportional: $\hat{P}_{0, \alpha}^{AB} = \delta \hat{Q}_B$;

\item
 $\hat{P}_{1, \alpha}^{AB}$ is a linear combination of $\hat{P}_{AB}$ and $\gamma \mathbb{1}_A \otimes \hat{p}_B$. This last term vanishes upon setting the free parameter $\gamma = 0$ and the two spatial translation generators become equivalent: $\hat{P}_{1, \alpha}^{AB} = \hat{P}_{AB}$;

\item 
$\hat{K}_\alpha^{AB}$ is a linear combination of $\hat{D}_B$ and $\hat{K}_{AB}$: 
\begin{equation} \label{eq:K_alpha_AB}
	\hat{K}_\alpha^{AB} = \hat{K}_{AB} - \frac{1}{2} \frac{\kappa}{\hbar} \frac{m_B}{m_A} \hat{D}_B \ .
\end{equation}
\end{itemize}
Therefore, we have obtained that, at first order in $\alpha$, the exponents~\eqref{linearpsr} of the quantum Galilei group transformations are linear combination of five of the seven generators of the QRF transformations~\eqref{eq:dynamicalalgebra} encountered in~\cite{Ballesteros:2020lgl}. In other words, at first order in $\alpha$, the quantum Galilei group element can be written in terms of the QRF transformations as
\begin{equation} \label{quantumGfirst}
    \begin{aligned}
        \hat{G}_\alpha &= e^{\frac{i}{\hbar} \hat{\theta} \otimes \hat{M}'} e^{\frac{i}{\hbar} \hat{b} \otimes \hat{P}_0'} e^{\frac{i}{\hbar} \hat{a} \otimes \hat{P}_1'} e^{\frac{i}{\hbar} \hat{v} \otimes \hat{K}'} \\
        &= e^{\frac{i}{\hbar} (- \frac{\kappa}{4 \hbar} \frac{m_B}{m_A} \hat{P}_{AB} + \frac{1}{2} \frac{m_B}{m_A} \hat{D}_A)} e^{\frac{i}{\hbar} \delta \hat{Q}_B} e^{\frac{i}{\hbar} \hat{P}_{AB}} e^{\frac{i}{\hbar} (\hat{K}_{AB} - \frac{\kappa}{2 \hbar} \frac{m_B}{m_A} \hat{D}_B)} \ .
    \end{aligned}
\end{equation}
The latter expression is an element of the Lie group generated by the seven dimensional QRF dynamical Lie algebra $\mathcal{D}(7)$~\eqref{eq:dynamicalalgebra}. In fact, it is straightforward to check that the four operators~\eqref{qgelementsansatz} in the exponents do not close a four-dimensional Lie algebra\footnote{Note that the group property~\eqref{gp} only ensures that the exponents close a four-dimensional Lie algebra when the coordinates $(a, v, b, \theta)$ are commutative objects, as in the case of Lie groups. In the case of quantum groups the exponents appearing in the group element generate in general an infinite-dimensional algebra. This is indeed the case if we consider the quantum Galilei group exponents~\eqref{expT} at all orders in $\alpha$. The fact that at first order in $\alpha$ the exponents and their commutators generate a 7-dim Lie algebra is quite an exceptional fact, that can be explained because in this case the phase space realisation of the quantum group exponents consists in at most quadratic functions of the phase space operators.}, but when their commutators are considered as additional generators of a higher dimensional structure, the latter turns out to be the seven dimensional QRF dynamical Lie algebra~\eqref{eq:dynamicalalgebra}. Alternatively, this fact can be also proven by realising that
$[\hat{K}_{AB}, \hat{D}_B]$ contains $\hat{T}$ and $[\hat{K}_{AB}, \hat{T}]$ contains $\hat{Q}_A$, and therefore both $\hat{T}$ and $\hat{Q}_A$ will arise within the commutators that come out when $e^{\frac{i}{\hbar}(\hat{K}_{AB} - \frac{1}{2} \frac{\kappa}{\hbar} \frac{m_B}{m_A} \hat{D}_B)}$ is written as the product of uniparametric subgroups. 
 
Therefore, closing the algebra defined by the operators~\eqref{qgelementsansatz} makes the correspondence between the quantum Galilei operators and the QRF operators complete. The four operators $(\hat{M}_\alpha^{AB}, \hat{P}_{0, \alpha}^{AB}, \hat{P}_{1, \alpha}^{AB}, \hat{K}_\alpha^{AB})$ generate through commutations the full 7-dimensional dynamical Lie algebra $\mathcal{D}(7)$ and the complete algebraic equivalence between the quantum Galilei group approach here presented and the dynamical Lie group introduced in~\cite{Ballesteros:2020lgl} is demonstrated. This is what establishes the physical interpretation of the quantum Galilei group as describing QRF symmetries, since the exponents of the quantum Galilei group generate exactly the same group of symmetries as the QRF transformations.
 
We stress that this equivalence holds only at first order in the parameter $\alpha$ which deforms the algebra of quantum group coordinates and of quantum Galilei generators. In accordance with the physical interpretation of the quantum Galilei group as QRF symmetries, this same parameter $\alpha$ is identified with the inverse of the mass of the particle which defines the reference frame in the QRF setting. Indeed, when $m_A \to \infty$, which in the QRF context brings us back to the idealised, nondynamical reference frame limit,  the quantum deformation parameter $\alpha \to 0$, so that one recovers the standard Galilei group. Finally, the variable $\hat{p}_A'$ which, in the QRF approach, corresponds to the momentum $\hat{\pi}_A$ of the particle is a ``renormalisation'' of the ``natural'' phase space variable $\hat{p}_A$ which is used to represent the quantum group coordinates on a phase space.  

For completeness, we also write the action of the quantum Galilei group exponentials\footnote{The action of a generic operator $\hat{U}^\alpha \equiv e^X$ onto a phase space coordinate $Y$ can be achieved by making use of the well-known identity $e^{X}Ye^{-X} = \sum_{n = 0}^{\infty} {\frac {\mbox{Ad}^n_X(Y)}{n!}}$, where $\mbox{Ad}^n_X(Y) \equiv \underbrace{[X, \cdots [X, [X}_{n{\text{ times }}}, Y]] \cdots] \ , \quad [(X)^{0}, Y] \equiv Y$.} \eqref{eq:Qgroupele} on the variables $(\hat{q}_A, \hat{p}_A') \otimes (\hat{q}_B, \hat{p}_B)$, with  $[\hat{q}_B, \hat{p}_B] = i \hbar$ and $[\hat{q}_A, \hat{p}_A'] = i \kappa$:\footnote{Notice that when working at first order in $\alpha$ and taking the ansatz $\alpha = (m_A \phi)^{-1}$, both momentum variables $\hat{p}_A$ and $\hat{p}_A'$ are conjugate to the variable $\hat{q}_A$ and satisfy the same commutation relation with $\hat{q}_A$.}
\begin{equation}
    \begin{aligned}
        \hat{U}_M^\alpha \, \hat{q}_A \, (\hat{U}_M^\alpha)^{-1} &= \left(1 + \frac{1}{2} \frac{\kappa}{\hbar} \frac{m_B}{m_A}\right) \hat{q}_A \ , \\
        \hat{U}_M^\alpha \, \hat{p}_A' \, (\hat{U}_M^\alpha)^{-1} &= \left(1 - \frac{1}{2} \frac{\kappa}{\hbar} \frac{m_B}{m_A}\right) \hat{p}_A' + \frac{1}{4} \frac{\kappa^2}{\hbar^2} \frac{m_B}{m_A} \hat{p}_B + \frac{\phi}{8} \frac{\kappa^2}{\hbar^2} \frac{m_B^2}{m_A} \ , \\
        \hat{U}_M^\alpha \, \hat{q}_B \, (\hat{U}_M^\alpha)^{-1} &= \hat{q}_B - \frac{1}{4} \frac{\kappa}{\hbar} \frac{m_B}{m_A} \hat{q}_A  \ , \\
        \hat{U}_M^\alpha \, \hat{p}_B \, (\hat{U}_M^\alpha)^{-1} &= \hat{p}_B \ ;
    \end{aligned}
\end{equation}
\begin{equation}
    \begin{aligned}
        \hat{U}_{P_0}^\alpha \, \hat{q}_A \, (\hat{U}_{P_0}^\alpha)^{-1} &= \hat{q}_A \ , &\qquad \hat{U}_{P_0}^\alpha \, \hat{p}_A' \, \hat{U}_{P_0}^\alpha)^{-1} &= \hat{p}_A' \ , \\
        \hat{U}_{P_0}^\alpha \, \hat{q}_B \, (\hat{U}_{P_0}^\alpha)^{-1} &= \hat{q}_B + \frac{\delta}{m_B} \hat{p}_B \ , &\qquad \hat{U}_{P_0}^\alpha \, \hat{p}_B \, (\hat{U}_{P_0}^\alpha)^{-1} &= \hat{p}_B \ ;
    \end{aligned}
\end{equation}
\begin{equation}
    \begin{aligned}
        \hat{U}_{P_1}^\alpha \, \hat{q}_A \, (\hat{U}_{P_1}^\alpha)^{-1} &= \hat{q}_A \ , &\qquad \hat{U}_{P_1}^\alpha \, \hat{p}_A' \, (\hat{U}_{P_1}^\alpha)^{-1} &= \hat{p}_A' - \frac{\kappa}{\hbar} \hat{p}_B \ , \\
        \hat{U}_{P_1}^\alpha \, \hat{q}_B \, (\hat{U}_{P_1}^\alpha)^{-1} &= \hat{q}_A + \hat{q}_B + \gamma \ , &\qquad \hat{U}_{P_1}^\alpha \, \hat{p}_B \, (\hat{U}_{P_1}^\alpha)^{-1} &= \hat{p}_B \ ;
    \end{aligned}
\end{equation}
\begin{equation}
    \begin{aligned}
        \hat{U}_K^\alpha \, \hat{q}_A \, (\hat{U}_K^\alpha)^{-1} &= \hat{q}_A 
        - \frac{\kappa}{\hbar} \frac{1}{m_A} (m_B \hat{q}_B - t \hat{p}_B) \ , \\
        \hat{U}_K^\alpha \, \hat{p}_A' \, (\hat{U}_K^\alpha)^{-1} &= \hat{p}_A' \ , \\
        \hat{U}_K^\alpha \, \hat{q}_B \, (\hat{U}_K^\alpha)^{-1} &= 
        \left(1 - \frac{1}{2} \frac{\kappa}{\hbar} \frac{m_B}{m_A}\right) \hat{q}_B + \frac{{p}_A'}{m_A} t - \frac{\phi}{4} \frac{\kappa}{\hbar} \frac{m_B}{m_A} t \ , \\
        \hat{U}_K^\alpha \, \hat{p}_B \, (\hat{U}_K^\alpha)^{-1} &= 
        \left(1 + \frac{1}{2} \frac{\kappa}{\hbar} \frac{m_B}{m_A}\right) \hat{p}_B + \frac{m_B}{m_A} \hat{p}_A' + \frac{\phi}{4} \frac{\kappa}{\hbar} \frac{m_B^2}{m_A} \ .
    \end{aligned}
\end{equation}


\sect{Poisson--Lie group limit and dynamical reference frames}
\label{sec:PLgrouplimit}

In the context of QRFs it is possible to consider the limit in which the $A$ particle serving as the reference frame is classical. In this case one has a dynamical reference frame (see discussion in \cite{Ballesteros:2020lgl}), where the Poisson brackets of the phase space variables are deduced from the commutator~\eqref{commkh} via the usual limiting procedure:\footnote{Notice that we can take this limit separately for the particle $A$ thanks to the fact that we used for it a dedicated noncommutativity parameter, $\kappa$.}
\begin{equation} \label{Poisson}
    \{x_A, p_A\} = \lim_{\kappa \to 0} \frac{[\hat{x}_A, \hat{p}_A]}{i \kappa} = 1 \ .
\end{equation}
Given the correspondence between QRF and quantum Galilei groups that we have exposed in the previous Sections, it is natural to ask what structures are generated when performing an analogous classical limit in the quantum Galilei group setting. We first discuss this on a formal algebraic level, and in the next Section we provide more physical insights.

As one could have easily guessed at this point, by applying the same limiting procedure as in~\eqref{Poisson} to the algebra of quantum group coordinates~\eqref{qga},
\begin{equation}
    [\hat{a}, \hat{v}] = i \kappa \alpha \hat{v} \ , \qquad [\hat{a}, \hat{\theta}] = i \kappa \alpha \hat{\theta} \ , \qquad [\hat{v}, \hat{\theta}] = - \frac{i}{2} \kappa \alpha \hat{v}^2 \ , \qquad [\hat{b}, \cdot] = 0 \ ,
\end{equation}
we obtain the Poisson algebra of commutative Galilei coordinates,
\begin{equation} \label{pllimit}
    \{a, v\} = \alpha v \ , \qquad \{a, \theta\} = \alpha \theta \ , \qquad \{v, \theta\} = - \frac{1}{2} \alpha v^2 \ , \qquad \{b, \cdot\} = 0 \ ,
\end{equation}
which was already described in Subsection~\ref{sec:PLgroup} and defines the Poisson--Lie structure on the extended Galilei group that underlies the full construction. In fact, the symplectic leaves of the Poisson--Lie structure~\eqref{pllimit} are given by the Casimir functions
\begin{equation}
    \gamma = a - 2 \frac{\theta}{v} \ , \qquad \delta = b \ .
\end{equation}
which provide the classical counterpart of the Casimir operators~\eqref{casimirsncc} and are again characterised by the parameters $(\gamma, \delta)$. Therefore, for each leaf, we can find a symplectic realisation of the Poisson structure~\eqref{plbracket1} in terms of one pair of canonical variables $(q_A, p_A)$ with canonical Poisson bracket $\{q_A, p_A\} = 1$, namely 
\begin{equation} \label{PLrealis}
    a = q_A + \gamma \ , \qquad v = \phi e^{\alpha p_A} \ , \qquad \theta = \frac{\phi}{2} q_A e^{\alpha p_A} \ , \qquad b = \delta \ ,
\end{equation}
whose quantum phase space analogue is just~\eqref{psncc} and the parameter $\phi$ has been again introduced by dimensional consistency. Note also that the parameter $\delta$ coincides with the commutative time coordinate $b$, and in this sense each time corresponds to a different symplectic leaf of the Poisson--Lie structure~\eqref{pllimit}.

Moreover, the same limit $\kappa \to 0$ transforms the commutators between the quantum Galilei algebra generators~\eqref{qalgqrfkappa} into those of the (undeformed) Galilei Lie algebra~\eqref{galileiphysical}, and gives the correct phase space realisation when applied to~\eqref{generatorsrepresentation}:
\begin{equation} \label{galileiundef}
    M' = m_B \ , \qquad P_0' = \frac{\hat{p}_B^2}{2 m_B} \ , \qquad  P_1' = \hat{p}_B \ , \qquad K' = - m_B \hat{q}_B + t \hat{p}_B \ .
\end{equation}
Summarising, the limit $\kappa \to 0$ brings us from the full quantum Galilei group phase space realisation~\eqref{expT} to the corresponding Poisson--Lie group realisation~\eqref{PLrealis} for particle $A$ and to the quantum phase space realisation~\eqref{galileiundef} of the (undeformed) Galilei Lie algebra for particle $B$. Therefore, the exponents of the quantum Galilei group transformations~\eqref{eq:Qgroupele} become a product of classical phase space functions for particle $A$ and quantum phase space operators for particle $B$:
\begin{equation} \label{exponentsk0}
    \begin{aligned}
        \hat{M}_c^{AB} &= \frac{\phi}{2} q_A e^{\alpha p_A} \otimes m_B \mathbb{1}_B \ , \\
        \hat{P}_{0, c}^{AB} &= \delta \otimes \frac{\hat{p}_B^2}{2 m_B} \ , \\
        \hat{P}_{1, c}^{AB} &= (q_A + \gamma) \otimes \hat{p}_B \ , \\
        \hat{K}_c^{AB} &= \phi e^{\alpha p_A} \otimes (t \hat{p}_B - m_B \hat{q}_B) \ .
    \end{aligned}
\end{equation}
Note that if we now expand the previous equations at first order in $\alpha$, we get 
\begin{equation} \label{eq:McPLgroup}
    \begin{aligned}
        \hat{M}_c^{AB} &= \frac{\phi}{2} q_A (1 + \alpha p_A) \otimes m_B \mathbb{1}_B = q_A \frac{p_A'}{2 m_A} \otimes m_B \mathbb{1}_B \ , \\
        \hat{P}_{0, c}^{AB} &= \delta \otimes \frac{\hat{p}_B^2}{2 m_B} \ , \\
        \hat{P}_{1, c}^{AB} &= q_A \otimes \hat{p}_B \ , \\
        \hat{K}_c^{AB} &= \phi (1 + \alpha p_A) \otimes (t \hat{p}_B - m_B \hat{q}_B) = \frac{p_A'}{m_A} \otimes (t \hat{p}_B - m_B \hat{q}_B) \ ,
    \end{aligned}
\end{equation}
where we used the ansatz $\alpha = (m_A \phi)^{-1}$ and we set $\gamma = \xi = 0$.


\section{QRF transformations on superpositions of semiclassical states}
\label{sec:QGGasQRF}

We have so far demonstrated that the limit $\kappa \to 0$ connects the quantum Galilei group to its corresponding Poisson--Lie group, both at the formal algebraic level  and at the level of the phase space realisation.
However, this connection is valid on an even deeper level, related to the physical interpretation of the quantum Galilei group (and its Poisson--Lie counterpart) as QRF transformations.

The quantum Galilei group acts on the Hilbert space $\mathcal{H} = \mathcal{H}_A \otimes \mathcal{H}_B$, which is the product of the Hilbert spaces with phase space variables $(\hat{q}_A, \hat{p}_A')$ and $(\hat{q}_B, \hat{p}_B)$ respectively. Quantum states living in $\mathcal{H}$ are given by (combinations of)  products of the kind $|\phi\rangle_A \, |\psi\rangle_B$: given the algebraic correspondence between the quantum Galilei group (at first order in $\alpha$) and the QRF transformations, we can think about $|\phi\rangle_A$ and $|\psi\rangle_B$ as the states of the quantum particles $A$ and $B$. 
In particular, we can take semiclassical states in the Hilbert space $\mathcal{H}_A$, $\ket{\chi(t)}_A$, where by semiclassical we mean a quantum state that evolves according to some Hamiltonian operator $\hat{H}_A$ and that, for a time interval $t \in [t_i, t_f]$ in which we consider the time evolution, approximately satisfies the conditions
\begin{equation}\label{semiclassicalcondition}
    \hat{q}_A \ket{\chi(t)}_A \approx q_A(t) \ket{\chi(t)}_A \ ; \qquad \hat{p}_A' \ket{\chi(t)}_A \approx p_A'(t) \ket{\chi(t)}_A \ ,
\end{equation}
where $(q_A(t), p_A'(t))$ is the solution of the classical equations of motion for the corresponding classical Hamiltonian $H_A$.

The semiclassical condition implies that $[\hat{q}_A, \hat{p}'_A] \, \ket{\chi(t)}_A \approx 0$ when applied to the semiclassical state. Hence, it physically corresponds to taking the limit $\kappa \to 0$. As a consequence, we can consistently apply the $\kappa \to 0$ limit in Eq.~\eqref{Poisson} to the quantum group structures,   when we consider the action on the state $\ket{\chi(t)}_A \, |\psi\rangle_B$. 
Therefore, the Poisson--Lie group limit of the quantum group corresponds, in the language of QRFs, to a limit where the reference frame $A$ is described by a semiclassical (dynamical) state, while the other particle $B$ is fully quantum. We recall that, in the Poisson--Lie group limit, the algebra of the quantum group generators~\eqref{qalgqrfkappa} on $\mathcal{H}_B$ reduces to the usual Galilei group and the Poisson algebra of the transformation parameters on $A$ is deformed.

To see this limit even more explicitly, let us consider, for simplicity, the free-particle Hamiltonian $\hat{H}_A = \frac{\hat{p}_A'^2}{2 m_A}$. In this case, the state of particle $A$ is $\ket{\chi_f(t)}_A = e^{-\frac{i}{\kappa} \hat{H}_A t}\ket{\chi_f(0)}_A$, where $\ket{\chi_f(0)}_A$ is, e.g., a Gaussian wavefunction centred in $(q_A^0, p_A'^0)$. It is easy to see that
\begin{equation}
    \begin{aligned}
        \hat{q}_A \ket{\chi_f(t)}_A &\approx \left(q_A^0 + \frac{p_A'^0}{m_A} t\right) \ket{\chi_f(t)}_A \equiv q_A(t) \ket{\chi_f(t)}_A \ , \\
        \hat{p}_A' \ket{\chi_f(t)}_A &\approx p_A'^0 \ket{\chi_f(t)}_A \equiv p_A'(t) \ket{\chi_f(t)}_A \ .
    \end{aligned}
\end{equation}
Acting with the quantum group operators \eqref{qgelementsansatz} on the state $\ket{\chi_f(t)}_A \ket{\psi}_B$ we obtain
\begin{equation} \label{eq:GaussianPLgrouplimitG}
    \begin{aligned}
    	e^{\frac{i}{\hbar} \hat{M}_\alpha^{AB}} \ket{\chi(t)}_A \ket{\psi}_B &\approx e^{\frac{i}{\hbar}\frac{q_A(t) p_A'(t)}{2 m_A} \otimes \hat{M}} \ket{\chi(t)}_A \ket{\psi}_B \ , \\
        e^{\frac{i}{\hbar} \hat{P}_{0, \alpha}^{AB}} \ket{\chi(t)}_A \ket{\psi}_B &\approx e^{\frac{i}{\hbar} \delta \otimes \hat{P}_0} \ket{\chi(t)}_A \ket{\psi}_B \ , \\
        e^{\frac{i}{\hbar} \hat{P}_{1, \alpha}^{AB}} \ket{\chi(t)}_A \ket{\psi}_B &\approx e^{\frac{i}{\hbar} q_A(t) \otimes \hat{P}_1} \ket{\chi(t)}_A \ket{\psi}_B \ , \\
        e^{\frac{i}{\hbar} \hat{K}^{AB}_\alpha} \ket{\chi(t)}_A \ket{\psi}_B &\approx e^{\frac{i}{\hbar} \frac{p_A'(t)}{m_A} \otimes \hat{K}} \ket{\chi(t)}_A \ket{\psi}_B \ ,
    \end{aligned}
\end{equation}
where we accounted for the limit $\kappa \to 0$ and $(\hat{M}, \hat{P}_0, \hat{P}_1, \hat{K})$ are the generators of the undeformed Galilei group. As anticipated by the arguments exposed at the beginning of this Section, we immediately notice, comparing Eq.~\eqref{eq:GaussianPLgrouplimitG} to the expression of the Poisson--Lie group~\eqref{exponentsk0}-\eqref{eq:McPLgroup}, that the Poisson--Lie group arises in the limit in which the QRF has a semiclassical state and still retains a dynamical evolution. This is fully compatible with the QRF transformations of \cite{Ballesteros:2020lgl} that are obtained in the $\kappa \to 0$ limit.

Moreover, the formalism just developed allows us to understand how  QRF transformations emerge when going beyond semiclassical states for the reference frame $A$. Let us  consider a quantum superposition of two semiclassical states $\ket{\chi_1(t)}_A$ and $\ket{\chi_2(t)}_A$, centred, respectively, in $(q_1(t), p_1'(t))$ and $(q_2(t), p_2'(t))$, and such that $\braket{\chi_1(t)|\chi_2(t)} \approx 0$. Such a state is a quantum state. In this case, we can show that one needs to account for noncommutative transformation parameters, and that QRF transformations are recovered as a limit of the quantum group transformations. 

Let us consider, for instance, the quantum Galilei boost transformation. In this case, we can effectively write
\begin{equation}
	e^{\frac{i}{\hbar} \hat{K}^{AB}_\alpha} \frac{1}{\sqrt{2}} (\ket{\chi_1(t)}_A + \ket{\chi_2(t)}_A) \ket{\psi(t)}_B \approx \frac{1}{\sqrt{2}} \left(\ket{\chi_1(t)}_A e^{\frac{i}{\hbar} \frac{p_{A, 1}'(t)}{m_A} \otimes \hat{K}} + \ket{\chi_2(t)}_A e^{\frac{i}{\hbar} \frac{p_{A, 2}'(t)}{m_A} \otimes \hat{K}}\right) \ket{\psi(t)}_B \ ,
\end{equation}
where we used Eq.~\eqref{eq:GaussianPLgrouplimitG} to write the action on the individual semiclassical states $\ket{\chi_i(t)}_A$, $i = 1, 2$, obtaining a ``controlled'' quantum superposition of standard Galilei transformations. Here we clearly see that the linear superposition of two semiclassical states is not a semiclassical state: we cannot use a single dynamical variable $p_A'(t)$ to describe the action of the operator $\hat{K}_\alpha^{AB}$ on the state $\frac{1}{\sqrt{2}} (\ket{\chi_1(t)}_A + \ket{\chi_2(t)}_A) \ket{\psi(t)}_B$.
We can however use the semiclassical condition~\eqref{semiclassicalcondition} and write, effectively,
\begin{equation}
	p_{A, i}'(t) \ket{\chi_i(t)}_A \approx \hat{p}_A' \ket{\chi_i(t)}_A = \hat{\pi}_A \ket{\chi_i(t)}_A \ ,
\end{equation}
where in the last step we made use of the identification of the quantum Galilei group phase space operator $\hat{p}_A'$ with the QRF phase space variable of particle $A$, $\hat{\pi}_A$. This allows us to show that the ``controlled'' quantum superposition of standard Galilei transformations is in fact a QRF transformation:
\begin{equation}
	\frac{1}{\sqrt{2}} \left(\ket{\chi_1(t)}_A e^{\frac{i}{\hbar} \frac{p_1'(t)}{m_A} \otimes \hat{K}} + \ket{\chi_2(t)}_A e^{\frac{i}{\hbar} \frac{p_2'(t)}{m_A} \otimes \hat{K}}\right) \ket{\psi(t)}_B \approx e^{\frac{i}{\hbar} \frac{\hat{\pi}_A}{m_A} \otimes \hat{K}} \frac{1}{\sqrt{2}} \left(\ket{\chi_1(t)}_A + \ket{\chi_2(t)}_A\right) \ket{\psi(t)}_B \ .
\end{equation}
In other words, the action of a quantum Galilei boost on the linear superposition of semiclassical states is equivalent to the action of the operator associated to a QRF boost that was defined in \cite{Giacomini:2017zju}.  This equivalence only holds for quantum states that can be written as superposition of semiclassical states: for generic quantum states we have to resort to the full quantum Galilei boost operator $\hat{K}_\alpha^{AB}$, which, at first order in $\alpha$, is a linear combination of the QRF boost and the dilation $\hat{D}_B$ (see Subsection~\ref{sec:correspondence}). Analogue reasoning works for the other elements of the quantum Galilei group.

This clarifies that  superpositions of Galilei transformations, which were used in \cite{Giacomini:2017zju} to construct QRF transformations, are physically relevant when the state of the QRF is in a quantum superposition of semiclassical states. In other words, these transformations define a symmetry of QRFs (described by the Lie group of inertial QRF transformations $\mathcal{D}(7)$ in~\cite{Ballesteros:2020lgl}) when the quantum state of the QRFs are restricted to superpositions of (approximately) orthogonal  semiclassical states. In this work, we showed that the same algebra of symmetries $\mathcal{D}(7)$ can be described by the first-order deformation of the Galilei group into a quantum group. We conjecture that if one wants to consider arbitrary quantum states of the QRF, then  the symmetry group acquires nontrivial features and becomes a full-fledged quantum Galilei group (at all orders in $\alpha$) under the form of the phase space realisation of its associated universal T-matrix, which is given in Eq.~\eqref{expT}. 


\sect{Concluding remarks and open questions} \label{sec:concluding}

The correspondence between QRF transformations and quantum groups that we have uncovered sheds light on the physical relevance of each of the two formalisms.

On the one hand, we showed that quantum groups have a physical realisation as transformations between quantum reference frames:   noncommutative transformation parameters are operators acting on the phase space of a quantum system that plays the role of a quantum reference frame, and quantum Galilei generators are operators acting on the phase space of the quantum system that is being transformed. The parameter governing the quantum deformation is inversely proportional to the mass of the quantum reference frame. To the order of the deformation parameter considered here, the other parameter $\phi$ should correspond to a velocity scale of the particle serving as the QRF, but in fact it does not seem to play a physical role, because it can always be incorporated in a redefinition of the physical variables.  

On the other hand, we showed that we can interpret the QRF transformations developed in \cite{Giacomini:2017zju} as describing some intermediate regime between having classical dynamical transformations and transformations between generic quantum systems.  Specifically, the QRF regime of~\cite{Giacomini:2017zju}, can be described by the first-order deformation of the standard Galilei Lie group into a quantum group and applies to physical systems that are in a superposition of semiclassical states. In fact, if we act with quantum Galilei group operators (taken at first order in the quantum deformation parameter) onto superpositions of semiclassical states for the QRF, we recover the ``controlled'' action on these states that we would obtain from the superpositions of boosts and translations given in~\cite{Giacomini:2017zju}. The Poisson--Lie generalisation of the Galilei group (obtained from the quantum group in the limit $\kappa \to 0$) describes dynamical reference frames transformations, consistently with the fact that in the $\kappa \to 0$ limit the quantum phase space of the particle playing the role of reference frame becomes a classical phase space. Note that in this limit, while the quantum phase space parameter $\kappa$ of the QRF particle vanishes, the one of the other particle, $\hbar$, is nonvanishing. Therefore the Poisson--Lie Galilei group acts on the product of a classical phase space and a noncommutative quantum phase space. 

Moreover, thanks to our construction, the symmetry properties of QRF transformations are understood as deriving from the properties of quantum group transformations. This solves a long-standing issue, related to the observation that the operators describing the superposition of translations and boosts in the context of QRF proposed in~\cite{Giacomini:2017zju} do not close a group of transformations. In~\cite{Ballesteros:2020lgl} we had found that the generators of the transformations arising when composing the QRF generalisation of Galilei translations and boosts given in~\cite{Giacomini:2017zju} close a seven-dimensional Lie algebra, $\mathcal{D}(7)$. Here we go one step further: we show that $\mathcal{D}(7)$ results from taking the first order in the quantum deformation parameter of the operators generating the quantum Galilei group written under a T-matrix presentation. These by definition generate a quantum group, which is an infinite-dimensional object, but, at first order in the quantum deformation parameter, we obtain the finite-dimensional algebra $\mathcal{D}(7)$. 

Summarising, the results and observations we just exposed lead us to conjecture that all-order quantum Galilei group operators provide the appropriate generalisation of QRF transformations acting on arbitrary quantum states. 

It would be interesting to explore whether the quantum Galilei group associated to QRF symmetries leads to the existence of a minimal localisation of a quantum particle, thus effectively realising a discrete ``quantum space'' structure. This could be achieved, for instance, using an approach based on pregeometry as considered in~\cite{Amelino-Camelia:2012vlk}.

Finally, the quantum group approach here presented could be generalised to higher dimensions, thus paving the way to (2+1)- and (3+1)-dimensional generalisations of the QRF transformations. This motivates the construction of the (2+1) and (3+1) analogues of the quantum Galilei group here presented, including their corresponding universal T-matrices. It is worth stressing that, to the best of our knowledge, the only existing (3+1)-dimensional centrally extended quantum Galilei group was constructed in~\cite{deAzcarraga:1996mij}, but that is not the type of quantum deformation considered in this paper. Another challenging problem would consist in finding the quantum Poincar\'e group whose nonrelativistic limit gives rise to the quantum Galilei group here described. Should this be achieved, it would open the path to the definition of a relativistic generalisation of QRFs endowed with quantum Poincar\'e group invariance. Work along all these lines is currently in progress.


\section*{Acknowledgements}

A.B. and D.F-S. acknowledge support from the grant PID2023-148373NB-I00 funded by MCIN/AEI/ 10.13039/501100011033/FEDER -- UE, and the Q-CAYLE Project funded by the Regional Government of Castilla y Le\'on (Junta de Castilla y Le\'on) and the Ministry of Science and Innovation MICIN through NextGenerationEU (PRTR C17.I1). D.F-S. acknowledges support from Universidad de Burgos through a PhD grant. F.G. acknowledges support from the Swiss National Science Foundation via the Ambizione Grant PZ00P2-208885, from the ETH Zurich Quantum Center, and from the ID\#~62312 grant from the John Templeton Foundation, as part of the `The Quantum Information Structure of Spacetime, Second Phase (QISS 2)' Project. The opinions expressed in this publication are those of the authors and do not necessarily reflect the views of the John Templeton Foundation. This work contributes to the COST Action CA23130 ``Bridging high and low energies in search of quantum gravity (BridgeQG)''.


\appA{Appendix. The Galilei universal T-matrix} \label{appendix1}

\setcounter{equation}{0}
\renewcommand{\theequation}{A.\arabic{equation}}

The concept of universal T-matrix (or exponential map for quantum groups) arises whenever we have two mutually dual Hopf algebras~\cite{Fronsdal:1991gf, Fronsdal1994}. In our case, let us assume that we have a Hopf algebra of smooth functions on the quantum Galilei coordinates $(G_\alpha, \Delta_G)$ and its dual Hopf algebra of quantum Galilei algebra generators $(U_\alpha(\mathfrak{g}), \Delta_U)$, namely a quantum universal enveloping algebra. In the previous expressions, $G_\alpha$ and $U_\alpha(\mathfrak{g})$ denote the two algebras, and $\Delta_G$ and $\Delta_U$ denote the coproduct maps on each of them~\cite{Chari:1994pz}. If we further denote by $m_G$ and $m_U$ the noncommutative products in $G_\alpha$ and $U_\alpha(\mathfrak{g})$, respectively, the duality between the Hopf algebras  $(G_\alpha, m_G, \Delta_G)$ and $(U_\alpha(\mathfrak{g}), m_U, \Delta_U)$ is established through the existence of a canonical pairing $\langle \, , \, \rangle : G_\alpha \times U_\alpha(\mathfrak{g}) \rightarrow \mathbb{C}$ such that
\begin{equation} \label{pairing1}
	\langle m_G (f \otimes g), a \rangle = \langle f \otimes g, \Delta_U(a) \rangle \ ,
\end{equation}
\begin{equation} \label{pairing2}
	\langle \Delta_G(f), a \otimes b \rangle = \langle f, m_U (a \otimes b) \rangle \ ,
\end{equation}
where $a, b \in U_\alpha(\mathfrak{g})$, $f, g \in G_\alpha$, and $\langle f \otimes g, a \otimes b \rangle = \langle f, a \rangle \, \langle g, b \rangle.$ Essentially, the duality between $(G_\alpha, m_G, \Delta_G)$ and $(U_\alpha(\mathfrak{g}), m_U, \Delta_U)$ means that both noncommutative Hopf algebras contain the same algebraic information. This also happens in the commutative (undeformed) case, where the algebra of smooth functions on the Lie group $G$ and the universal enveloping algebra $U_\alpha(\mathfrak{g})$ of the Lie algebra $\mathfrak{g}$ of $G$ can be considered as dual objects under the same type of pairing. In this framework, the universal T-matrix is defined via the Hopf algebra dual form,
\begin{equation} \label{HADF}
	T = \sum_{\mu} \hat{z}_{\mu} \otimes \hat{Z}^{\mu} \ ,
\end{equation}
where $\hat{z}_\mu$ is an arbitrary element of the basis of the Hopf algebra $(G_\alpha, m_G, \Delta_G)$ and $\hat{Z}^\mu$ is an arbitrary element of the basis of the dual Hopf algebra $(U_\alpha(\mathfrak{g}), m_U, \Delta_U)$. 

In this Appendix, we explicitly construct the universal T-matrix of the quantum deformation of the centrally extended (1+1) Galilei group whose semiclassical limit is the Poisson--Lie structure~\eqref{plbracket1} with $\beta_1 = 0$. Therefore, we start from the noncommutative Hopf algebra of Galilei coordinates satisfying 
\begin{equation} \label{commrules}
	[\hat{a}, \hat{v}] = \alpha \hat{v} \ , \qquad
	[\hat{a}, \hat{\theta}] = \alpha \hat{\theta} \ , \qquad
	[\hat{v}, \hat{\theta}] = - \frac{1}{2} \alpha \hat{v}^2\ , \qquad
	[\hat{b}, \cdot] = 0 \ ,
\end{equation}
where the quantization parameter $\kappa$ has been omitted for the sake of simplicity, and whose coproduct map arises as the Hopf-algebraic transcription of the group law~\eqref{glaw}, namely
\begin{equation} \label{copmap}
	\begin{aligned}
    	\Delta_G(\hat{\theta}) &= \hat{\theta} \otimes 1 + 1 \otimes \hat{\theta} + \hat{v} \otimes \hat{a} + \frac{1}{2} \hat{v}^2 \otimes \hat{b} \ , \\
    	\Delta_G(\hat{b}) &= \hat{b} \otimes 1 + 1 \otimes \hat{b} \ , \\
    	\Delta_G(\hat{a}) &= \hat{a} \otimes 1 + 1 \otimes \hat{a}  + \hat{v} \otimes \hat{b} \ , \\
    	\Delta_G(\hat{v}) &= \hat{v} \otimes 1 + 1 \otimes \hat{v} \ .
	\end{aligned}
\end{equation}
It can be checked straightforwardly that this coproduct is an algebra homomorphism for the noncommutative algebra~\eqref{commrules}. The basis elements of $(G_\alpha, m_G, \Delta_G)$ are defined by the ordered monomials
\begin{equation}
	\hat{z}_{abcd} = \hat{\theta}^a \hat{b}^b \hat{a}^c \hat{v}^d \ ,
\end{equation}
where the exponents are nonnegative integers, and give the basis elements $\hat{Z}^{ijkl}$ of $(U_\alpha(\mathfrak{g}), m_U, \Delta_U)$ through the pairing
\begin{equation}
	\langle \hat{z}_{abcd}, \hat{Z}^{ijkl} \rangle = \delta_a^i \delta_b^j \delta_c^k \delta_d^l \ ,
\end{equation}
where
\begin{equation} \label{dualgenerators}
	\hat{Z}^{1000} = \hat{M} \ , \qquad
	\hat{Z}^{0100} = \hat{P}_0\ , \qquad
	\hat{Z}^{0010} = \hat{P}_1\ , \qquad
	\hat{Z}^{0001} = \hat{K} \ ,
\end{equation}
are the generators of the dual quantum Galilei algebra $(U_\alpha(\mathfrak{g}), m_U, \Delta_U)$. The explicit form of the basis elements $\hat{Z}^{ijkl}$ and their commutation rules, as well as the dual coproduct maps for the generators~\eqref{dualgenerators} have to be found by imposing the Hopf algebra duality conditions~\eqref{pairing1}-\eqref{pairing2}. The quantum Galilei algebra so obtained will be, by construction, the dual Hopf algebra to the one generated by the commutation rules~\eqref{commrules} and the coproduct~\eqref{copmap}.

More explicitly, the Hopf algebra of quantum coordinates has a coproduct and a noncommutative product defined generically as
\begin{equation}
	\begin{aligned}
    	\Delta_G(\hat{z}_{abcd}) &= E_{abcd}^{ijkl;pqrs} \,\hat{z}_{ijkl} \otimes \hat{z}_{pqrs} \ , \\
    	m_G(\hat{z}_{ijkl} \otimes \hat{z}_{pqrs}) &= F_{ijkl;pqrs}^{abcd} \, \hat{z}_{abcd} \ ,
	\end{aligned}
\end{equation}
so, by definition the coproduct and the product of the dual Hopf algebra read
\begin{equation}
	\begin{aligned}
		\Delta_U(\hat{Z}^{abcd}) &= F_{ijkl;pqrs}^{abcd} \hat{Z}^{ijkl} \otimes \hat{Z}^{pqrs} \ , \\
		m_U(\hat{Z}^{ijkl}\otimes \hat{Z}^{pqrs}) &= E_{abcd}^{ijkl;pqrs} \hat{Z}^{abcd} \ .
	\end{aligned}
\end{equation}
Notice that the duality is encoded in the role played by the structure tensors $E$ and $F$. One can easily check from~\eqref{copmap} that the structure tensor $E$ is such that
\begin{equation}
	\begin{aligned}
    	E_{abcd}^{0000;pqrs} &= \delta_a^p \delta_b^q \delta_c^r \delta_d^s\ , \\
    	E_{abcd}^{ijkl;0000} &= \delta_a^i \delta_b^j \delta_c^k \delta_d^l\ , \\
    	E_{0000}^{ijkl;pqrs} &= \delta_0^i \delta_0^j \delta_0^k \delta_0^l \delta_0^p \delta_0^q \delta_0^r \delta_0^s \ .
	\end{aligned}
\end{equation}
The relevant recurrence relations for the structure tensor $E$ can be extracted from the equations
\begin{equation}
	\begin{aligned}
    	\Delta_G(\hat{z}_{abcd}) &= \Delta_G(\hat{\theta}) \Delta_G(\hat{z}_{(a-1)bcd}) \ , \qquad (a \geq 1) \\
    	\Delta_G(\hat{z}_{abcd}) &= \Delta_G(\hat{b}) \Delta_G(\hat{z}_{a(b-1)cd}) \ , \qquad (b \geq 1) \\
    	\Delta_G(\hat{z}_{abcd}) &= \Delta_G(\hat{z}_{ab(c-1)d}) \Delta_G(\hat{a} + \alpha d) \ , \qquad (c \geq 1) \\
    	\Delta_G(\hat{z}_{abcd}) &= \Delta_G(\hat{z}_{abc(d-1)}) \Delta_G(\hat v) \ . \qquad (d \geq 1)
	\end{aligned}
\end{equation}
A lengthy computation gives rise to the following recurrence relations:
\begin{equation}
	\begin{aligned}
		E_{abcd}^{ijkl;pqrs} &= E_{(a-1)bcd}^{(i-1)jkl;pqrs} + E_{(a-1)bcd}^{ijkl;(p-1)qrs} \\
		&\quad + \sum_{i', k'} \binom{i'}{i} \binom{k'}{k} \frac{1}{2^{- i + i'}} (- \alpha)^{- i - k + i' + k'} (- i + i' + 1)^{- k + k'} \\
   		&\quad \cdot \left\lbrace\prod_{u = 0}^{- i + i' - 1} (u + 1) \left(E_{(a-1)bcd}^{i'jk'(i-i'+l-1);pq(r-1)s} + \alpha p E_{(a-1)bcd}^{i'jk'(i-i'+l-1);pqrs}\right) \right. \\
   		&\qquad + \left. \prod_{u = 0}^{- i + i' - 1} (u + 2) E_{(a-1)bcd}^{i'jk'(i-i'+l-2);p(q-1)rs}\right\rbrace \ , \qquad (a \geq 1) \\
   		E_{abcd}^{ijkl;pqrs} &= E_{a(b-1)cd}^{i(j-1)kl;pqrs} + E_{a(b-1)cd}^{ijkl;p(q-1)rs} \ , \qquad (b \geq 1) \\
   		E_{abcd}^{ijkl;pqrs} &= E_{ab(c-1)d}^{ij(k-1)l;pqrs} + E_{ab(c-1)d}^{ijkl;pq(r-1)s} + E_{ab(c-1)d}^{ijk(l-1);p(q-1)rs} + \alpha (d - l - s) E_{ab(c-1)d}^{ijkl;pqrs} \ , \qquad (c \geq 1) \\
   		E_{abcd}^{ijkl;pqrs} &= E_{abc(d-1)}^{ijk(l-1);pqrs} + E_{abc(d-1)}^{ijkl;pqr(s-1)} \ . \qquad (d \geq 1) \\
	\end{aligned}
\end{equation}
It can be proven that the above general expressions lead to the following components of the $E$ tensor:
\begin{equation}
	\begin{aligned}
    	E_{abcd}^{1000;pqrs} &= a \delta_a^{p+1} \delta_b^q \delta_c^r \delta_d^s \ , \qquad (a \geq 1) \\
    	E_{abcd}^{0100;pqrs} &= b \delta_a^p \delta_b^{q+1} \delta_c^r \delta_d^s \ , \qquad (b \geq 1) \\
    	E_{abcd}^{0010;pqrs} &= c \delta_a^p \delta_b^q \delta_c^{r+1} \delta_d^s \ , \qquad (c \geq 1) \\
    	E_{abcd}^{0001;pqrs} &= d \delta_a^p \delta_b^q \delta_c^r \delta_d^{s+1} \ . \qquad (d \geq 1)
	\end{aligned}
\end{equation}
Now, with this information, the dual basis $\hat{Z}^{ijkl}$ and its commutation rules can be derived by following standard procedures~\cite{Fronsdal:1991gf, Fronsdal1994, Bonechi:1993sn, Ballesteros1995, Ballesteros:1996awf}. In particular, one can show that
\begin{equation}
	\hat{Z}^{1000} \hat{Z}^{(p-1)qrs} = p \, \hat{Z}^{pqrs} \ ,
\end{equation}
hence
\begin{equation}
	\hat{Z}^{pqrs} = \frac{\hat{M}}{p} \hat{Z}^{(p-1)qrs} = \cdots = \frac{\hat{M}^p}{p!} \hat{Z}^{0qrs} \ .
\end{equation}
Analogous calculations show that the dual basis can be expressed in terms of the quantum Galilei algebra generators as
\begin{equation} \label{dualbasis}
	\hat{Z}^{pqrs} = \frac{\hat{M}^p}{p!} \frac{\hat{P}_0^q}{q!} \frac{\hat{P}_1^r}{r!} \frac{\hat{K}^s}{s!} \ .
\end{equation}
Therefore, the Galilei universal T-matrix can be finally constructed by substituting the two mutually dual basis into the Hopf algebra dual form~\eqref{HADF}, giving
\begin{equation} \label{finalT}
	T = \sum_{abcd} \hat{z}_{abcd} \otimes \hat{Z}^{abcd} = e^{\hat{\theta} \otimes \hat{M}} e^{\hat{b} \otimes \hat{P}_0} e^{\hat{a} \otimes \hat{P}_1} e^{\hat{v} \otimes \hat{K}} \ .
\end{equation}

Finally, the commutation relations between the quantum Galilei algebra generators follow from particular components of the structure tensor $E$ since, in general,
\begin{equation}
	[\hat{Z}^{ijkl}, \hat{Z}^{pqrs}] = (E_{abcd}^{ijkl;pqrs} - E_{abcd}^{pqrs;ijkl}) \, \hat{Z}^{abcd} \ .
\end{equation}
In particular, the following nontrivial commutation relations are obtained:
\begin{equation} \label{coalg}
	[\hat{P_0}, \hat{P_1}] = 0 \ , \qquad
	[\hat{K}, \hat{M}] = \frac{\alpha}{2} \hat{M}^2 e^{- \alpha \hat{P}_1} \ , \qquad
	[\hat{K}, \hat{P}_0] = \frac{1 - e^{- \alpha \hat{P}_1}}{\alpha} \ , \qquad
	[\hat{K}, \hat{P}_1] = \hat{M} e^{-\alpha \hat{P}_1} \ .
\end{equation}
The coproduct map for the generators of the quantum Galilei algebra follow from particular components of the structure tensor $F$. A closed expression for $F$ can be obtained by considering the product of arbitrary basis elements $\hat{z}_{abcd}$. Another cumbersome computation leads to
\begin{equation} \label{tensorF}
	\begin{aligned}
    	F_{ijkl;pqrs}^{abcd} &= \sum_{n = 0}^k \binom{k}{n} \binom{p}{a - i} \binom{r}{c - k + n} \frac{(- 1)^n}{2^{- a + i + p}} (- \alpha)^{- a - c + i + k + p + r} \\
    	&\quad \cdot (a - i)^n (d - s)^{- c + k + r - n} \prod_{u = 0}^{- a + i + p - 1} (l + u) \delta_{j + q}^b \delta_{i + l + p + s}^{a + d} \ .
	\end{aligned}
\end{equation}
Specifically, by taking into account that the quantum Galilei algebra generators are defined in the form~\eqref{dualgenerators} the following coproduct maps are obtained from~\eqref{tensorF}:
\begin{equation} \label{cogal}
	\begin{aligned}
    	\Delta_U(\hat{M}) &= \hat{M} \otimes 1 + e^{\alpha \hat{P}_1} \otimes \hat{M} \ , \\
    	\Delta_U(\hat{P}_0) &= \hat{P}_0 \otimes 1 + 1 \otimes \hat{P}_0 \ , \\
    	\Delta_U(\hat{P}_1) &= \hat{P}_1 \otimes 1 + 1 \otimes \hat{P}_1 \ , \\
    	\Delta_U(\hat{K}) &= \hat{K} \otimes e^{-\alpha \hat{P}_1} + 1 \otimes \hat{K} \ .
	\end{aligned}
\end{equation}
It can be straightforwardly checked that the quantum Galilei algebra defined by~\eqref{coalg} and~\eqref{cogal} is fully consistent with the quantisation of the Lie bialgebra structure given in~\cite{Ballesteros:1999ew}, which is the one associated to the Poisson--Lie Galilei group~\eqref{plbracket1} with $\beta_1 = 0$.

We stress that the limit $\alpha \to 0$ of the previous construction yields a commutative Hopf algebra of Galilei coordinates, with the same group law~\eqref{copmap}, whose dual Hopf algebra is the undeformed Galilei Lie algebra endowed with primitive coproducts, namely coproducts of the form $\Delta(Z) = Z \otimes 1 + 1 \otimes Z$. As a consequence, the $\alpha \to 0$ limit of the $T$-matrix~\eqref{dualbasis} is just the usual exponential map for the centrally extended Galilei Lie group in (1+1) dimensions.

It is also worth mentioning that an alternative derivation of the universal T-matrix~\eqref{finalT} was sketched in~\cite{Bonechi:1993sn} by using the quantum deformation of the Galilei Lie algebra of generators (in a different basis) as the initial Hopf algebra. We have obtained here the same T-matrix by considering instead the Hopf algebra generated by the quantum Galilei coordinates as initial data, and with all the computations described in more detail.


\bibliographystyle{utphys2}

\bibliography{biblio2}

\end{document}